\newif\ifpeerreview
\newif\ifdoublecolumn
\newif\ifsinglecolumn

\peerreviewfalse
\doublecolumntrue

\documentclass[a4paper,12pt,preprint]{elsarticle}

\usepackage{amssymb}

\journal{Magnetic Resonance Imaging.}

\usepackage{epstopdf} 

\usepackage{multirow}
\usepackage[cmex10]{amsmath}  
\usepackage{amssymb}  
\usepackage{float}
\usepackage{algorithmic}      
\usepackage[caption=false,font=footnotesize]{subfig} 
\usepackage{fixltx2e}
\usepackage{url} 
\usepackage{verbatim}
\usepackage{cases}  
\usepackage{gensymb} 
\usepackage{bm} 
\usepackage{color,soul} 

\usepackage{booktabs} 

\setulcolor{red} 
\setstcolor{green} 
\sethlcolor{yellow} 

\begin{document}

\begin{frontmatter}

\title{Iterative Self-consistent Parallel Magnetic Resonance Imaging Reconstruction based on Nonlocal Low-Rank Regularization \footnote{doi: 10.1016/\j.mri.2022.01.012.} \footnote{ \copyright $<2022>$. This manuscript version is made available under the CC-BY-NC-ND 4.0 license https://creativecommons.org/licenses/by-nc-nd/4.0/}}

\author[1]{Ting Pan}
\author[1]{Jizhong Duan\corref{cor1}}
\ead{ duanjz@kust.edu.cn}
\author[2]{Junfeng Wang}
\author[3]{Yu Liu}

\cortext[cor1]{Corresponding author}

\address[1]{Faculty of Information Engineering and Automation, Kunming University of Science and Technology, Kunming 650500, China.}
\address[2]{Department of Hepatobiliary Surgery, The First People’s Hospital of Yunnan Province, Kunming 650030, China.}
\address[3]{School of Microelectronics, Tianjin University, Tianjin 300072, China.}

\begin{abstract}
Iterative self-consistent parallel imaging reconstruction (SPIRiT) is an effective self-calibrated reconstruction model for parallel magnetic resonance imaging (PMRI). The joint L1 norm of wavelet or tight frame  coefficients and joint total variation (TV) regularization terms are incorporated into the SPIRiT model to improve the reconstruction performance. The simultaneous two-directional low-rankness (STDLR) in k-space data is incorporated into SPIRiT to realize improved reconstruction. Recent methods have exploited the nonlocal self-similarity (NSS) of images by imposing nonlocal low-rankness of similar patches to achieve a superior performance. To fully utilize both the NSS in Magnetic resonance (MR) images and calibration consistency in the k-space domain, we propose a nonlocal low-rank (NLR)-SPIRiT model by incorporating NLR regularization into the SPIRiT model. We apply the weighted nuclear norm (WNN) as a surrogate of the rank and employ the Nash equilibrium (NE) formulation and alternating direction method of multipliers (ADMM) to efficiently solve the NLR-SPIRiT model. The experimental results demonstrate the superior performance of NLR-SPIRiT over the state-of-the-art methods via three objective metrics and visual comparison.
\end{abstract}


\end{frontmatter}


\section{Introduction}
\label{introduction}

Magnetic resonance (MR) imaging (MRI) provides an indispensable imaging method without ionizing radiation in contemporary clinical applications. However, the scanning speed of MR imaging technique is limited. Both compressed sensing (CS) and parallel imaging (PI) techniques have been applied to reduce the MRI scanning time.

A variety of k-space undersampling patterns has been applied to reduce the amount of collected data, such as one-dimensional (1D) uniform undersampling (1DUU), 1D Gaussian random undersampling (1DGU), two-dimensional (2D) Poisson-disc undersampling (2DPU), 2D Gaussian random undersampling (2DGU), and non-Cartesian undersampling. According to the CS theory \cite{Candes2006, Lustig2007}, it is feasible to reconstruct the MR images from highly undersampled measurements \cite{Lustig2007}, since the MR images exhibit a sparsity in the wavelet transform domain and spatial finite differences. CS-MRI methods \cite{Lustig2007, Yang2010} have been proposed to solve reconstruction problems with regularization terms of the total variation (TV) and L1 norm of wavelet coefficients. Qu et al. developed the patch-based directional wavelet (PBDW and PBDWS) \cite{Qu2012, Ning2013} and graph-based redundant wavelet transform (GBRWT) \cite{Lai2016} to improve the reconstruction performance. In addition to fixed sparse transform methods, adaptive sparse representation-based reconstruction methods have been proposed, such as dictionary learning-based MRI (DLMRI) \cite{Ravishankar2011} and transform learning-based MRI (TLMRI) \cite{Ravishankar2015, Wen2015, Zhan2016, Ravishankar2017} algorithms. These algorithms have been proven to attain a high reconstruction performance.

Recently, some researchers have proposed certain methods to exploit the nonlocal self-similarity (NSS) of image patches to improve the image quality, such as the nonlocal means (NL-means) \cite{Manjon2010}, block-matching 3D denoising (BM3D) \cite{Dabov2007, Eksioglu2016}, patch-based nonlocal operator (PANO) \cite{Qu2014}, and nonlocal low-rank (NLR)-CS \cite{Dong2014} methods. The BM3D method exploits the NSS of image patches via their grouping for image denoising purposes \cite{Dabov2007, Eksioglu2016}, which has been applied in MRI reconstruction. Qu et al. \cite{Qu2014} exploited the NSS of image patches and established a PANO to reduce the reconstruction error. Dong et al. \cite{Dong2014} developed the very promising NLR-CS model employing NLR regularization of similar image patches constructed though block matching (BM).

Parallel MRI (PMRI) is also a well-known technique to accelerate MRI \cite{Pruessmann1999, Uecker2014, Lustig2010, Griswold2002}, which is often been combined with the CS theory to improve the reconstruction performance. Sensitivity encoding (SENSE) \cite{Pruessmann1999} explicitly utilizes sensitivity information. However, the performance of these methods is limited due to the difficulty in the accurate measurement of sensitivity information in practical applications. An iterative self-consistent parallel imaging reconstruction using eigenvector maps (ESPIRiT) model has been proposed  \cite{Uecker2014}. In contrast to the SENSE model, ESPIRiT model estimates multiple sets of coil sensitivity. We introduced the Lp pseudo-norm joint TV regularization term into the ESPIRiT scheme to improve the reconstruction performance \cite{Duan2019}.

Another type of PMRI reconstruction method avoids the difficulty in sensitivity estimation by implicitly considering sensitivity information\cite{Lustig2010}, such as the generalized autocalibrating partially parallel acquisitions (GRAPPA) \cite{Griswold2002, Chang2012} and iterative self-consistent parallel imaging reconstruction (SPIRiT) \cite{Lustig2009, Lustig2010, Murphy2012, Duan2014, Weller2014, Duan2018} methods, which relies on the k-space local kernel calibration. SPIRiT includes two reconstruction schemes in both the image \cite{Lustig2010, Murphy2012, Weller2014} and k-space \cite{Lustig2009, Lustig2010, Duan2014, Vasanawala2010} domains. An L1-SPIRiT scheme was obtained by combining the regularization term of the joint L1 norm (JL1) in the wavelet domain with the k-space domain-based SPIRiT model and solved with the projection over convex sets (POCS) algorithm \cite{Lustig2009, Duan2014}. Duan et al. \cite{Duan2018} applied the fast iterative shrinkage thresholding algorithm (FISTA) to solve the SPIRiT PMRI reconstruction problem in the k-space domain with the joint TV (JTV) regularization term. Weller et al. \cite{Weller2014} adopted the alternating direction method of multipliers (ADMM) technique to solve the SPIRiT PMRI reconstruction problem in the image domain with the JTV regularization term. The STDLR-SPIRiT scheme has been established \cite{Zhang2020} by combining the SPIRiT model with the simultaneous two-directional low-rankness (STDLR) in k-space to realize improved reconstruction. In fact, STDLR-SPIRiT has exploited the local low-rank (LR) prior rather than the LR feature based on nonlocal image structures. Most recently, Zhang et al. proposed the pFISTA-SPIRiT scheme \cite{Zhang2021} to solve the L1-SPIRiT model with tight frames, such as the shift-invariant discrete wavelets transform (SIDWT), and provided a guarantee convergence analysis.

Recent methods have exploited the NSS of images by imposing the group sparsity or the low-rankness of nonlocal similar patches to improve the reconstruction quality \cite{Manjon2010, Dabov2007, Eksioglu2016, Qu2014, Dong2014, Wen2020}. Among them, the NLR-based methods have achieved an excellent performance \cite{Dong2014, Wen2020}. In addition, there have been a variety of improved SPIRiT-based methods for PMRI reconstruction. However, to the best of our knowledge, no SPIRiT-based algorithm has applied low-rankness of nonlocal similar patches. In this paper, we propose an NLR-SPIRiT scheme incorporating NLR regularization into the SPIRiT model. The NLR-SPIRiT model fully utilizes both the NSS in MR images and the calibration consistency in the k-space domain to improve the reconstruction performance. By employing the Nash equilibrium (NE) formulation \cite{Danielyan2012, Yoon2014}, we reformulate the NLR-SPIRiT model into a two-objective optimization problem including a rank minimization problem and a least-squares (LS) problem. We adopt the weighted nuclear norm (WNN) \cite{Gu2014, Lu2016} instead of the nuclear norm (NN) as a surrogate of the rank, which yields a more efficient method to solve the rank minimization problem. The LS problem is efficiently solved with the ADMM technique \cite{Afonso2010}. The experimental results demonstrate the superior performance of the NLR-SPIRiT model over state-of-the-art methods in terms of three objective metrics and visual comparison.

We organize the rest of this article as follows: in Section \ref{overview}, we review the SPIRiT model and SPIRiT-based algorithms. In Section \ref{algorithm}, we describe the NLR-SPIRiT model, which incorporates NLR regularization of similar patches into the SPIRiT model. We adopt the WNN as a surrogate of the rank and employ the NE formulation and ADMM technique to efficiently solve the NLR-SPIRiT model. Section \ref{result} presents the experimental results, analysis, and discussion. Finally, we provide the conclusion of this paper in Section \ref{conclusion}.

\section{Related work}
\label{overview}

\subsection{Overview of the nonlocal low-rank penalty problem}

Suppose $\hat X \in {\mathbb{C}^N}$  represents a single coil image stacked in column, and $\hat Y \in {\mathbb{C}^M}$ represents the undersampled k-space data of the single coil image stacked in column. $P \in {\mathbb{R}^{M \times N}}$ is an undersampled operator selecting only the acquired k-space data from the entire k-space grid, ${U_y} \in {\mathbb{C}^{{N_y} \times {N_y}}}$ and ${U_x} \in {\mathbb{C}^{{N_x} \times {N_x}}}$  are discrete 2D Fourier transform matrices, $F = {U_y} \otimes {U_x} \in {\mathbb{C}^{N \times N}}$ is the Fourier operator applied on the single coil image. $\hat X$ is divided into ${N_p}$ overlapping patches. The mapping ${V_i}:\hat X \mapsto {V_i}(\hat X)$ is a BM operator, where ${V_i}(\hat X)$ represents the similar patch group matrix of the ${i^{th}}$ reference patch \cite{Dong2014}. Because all ${V_i}(\hat X)$ are LR matrices, the patch-based self-similarity constraint can be written as the following NLR penalty:
\begin{equation}
\Psi (\hat X) = \sum\limits_{i = 1}^{{N_p}} {\operatorname{rank} ({{V_i}(\hat X)})} \label{eq01}
\end{equation}
where  ${\operatorname{rank} ({{V_i}(\hat X)})}$ denotes the rank of the matrix  ${{V_i}(\hat X)}$. With the NLR penalty, reconstructing a single coil image $\hat X$  from k-space undersampled data $\hat Y$  can be usually reformulated as the following minimization problem:
\begin{equation}
\hat X = \arg \;\mathop {\min }\limits_{\hat X} \frac{1}{2}\left\| {PF\hat X - \hat Y} \right\|_2^2 + \alpha \Psi (\hat X) \label{eq02}
\end{equation}

\subsection{Overview of the Nash equilibrium formulation}

Considering an optimization problem regarding variables $p$ and $q$, as follows:
\begin{equation}
(p,q) = \arg \;\mathop {\min }\limits_{p,q} {f_1}(p,q) + {f_2}(p) + {f_3}(q) \label{eq03}
\end{equation}
where  ${f_1}(p,q)$ is the function of $p$ and $q$, ${f_2}(p)$ the function of $p$, and ${f_3}(q)$ is the function of $q$. Employing the Nash equilibrium (NE) formulation \cite{Danielyan2012, Yoon2014}, problem (\ref{eq03}) can be transformed to the following two problems:
\begin{equation}
\left\{ \begin{gathered}
  {p^ * } = \arg \;\mathop {\min }\limits_p {L_1}(p) \hfill \\
  {q^ * } = \arg \;\mathop {\min }\limits_q {L_2}(q) \hfill \\
\end{gathered}  \right. \label{eq04}
\end{equation}
where:
\begin{equation}
{L_1}(p) = {f_1}(p,{q^ * }) + {f_2}(p) \label{eq05}
\end{equation}
\begin{equation}
{L_2}(q) = {f_1}({p^ * },q) + {f_3}(q) \label{eq06}
\end{equation}
where ${p^*}$ and ${q^*}$ are both fixed values. According to the game theory, problem (\ref{eq03}) can be interpreted as a game between $p$ and $q$. Minimizing  ${L_1}(p)$ will increase the value of ${L_2}(q)$ and minimizing ${L_2}(q)$ will increase ${L_1}(p)$. The equilibrium of this game is called Nash equilibrium, which finds a balance between $p$ and $q$. Especially, problem (\ref{eq03}) is optimal in the fixed point $({p^ * },{q^ * })$.

\subsection{Overview of the SPIRiT model}

 Suppose $X \in \mathbb{C}{^{NC}}$ represents a multicoil image stacked in columns, $Y \in \mathbb{C}{^{MC}}$ denotes the undersampled k-space data of the multicoil image stacked in columns, ${\cal F} = {I_C} \otimes F \in \mathbb{C}{^{NC \times NC}}$, ${\cal P} = {I_C} \otimes P \in \mathbb{C}{^{MC \times NC}}$, $A = {\cal P}{\cal F} \in \mathbb{C}{^{MC \times NC}}$ represents the undersampled encoding matrix, ${I_C}$ is a $C \times C$ identity matrix, $ \otimes $ denotes Kronecker product, $N = {N_x} \times {N_y}$, where ${N_x}$ and ${N_y}$ are the number of rows and columns of the single-coil image, respectively, and $C$ is the total number of coils. The undersampled k-space data of multicoil images are thus given by:
\begin{equation}
Y = AX \label{eq1}
\end{equation}

\begin{figure*}[!t]
\footnotesize{
\centering{ %

\newcommand{\uwidth}{5.5}   
\newcommand{\uhoriz}{-0.5}  

\begin{minipage}[b]{ 0.976 \textwidth} %
\centering{
\includegraphics[width= \uwidth in]{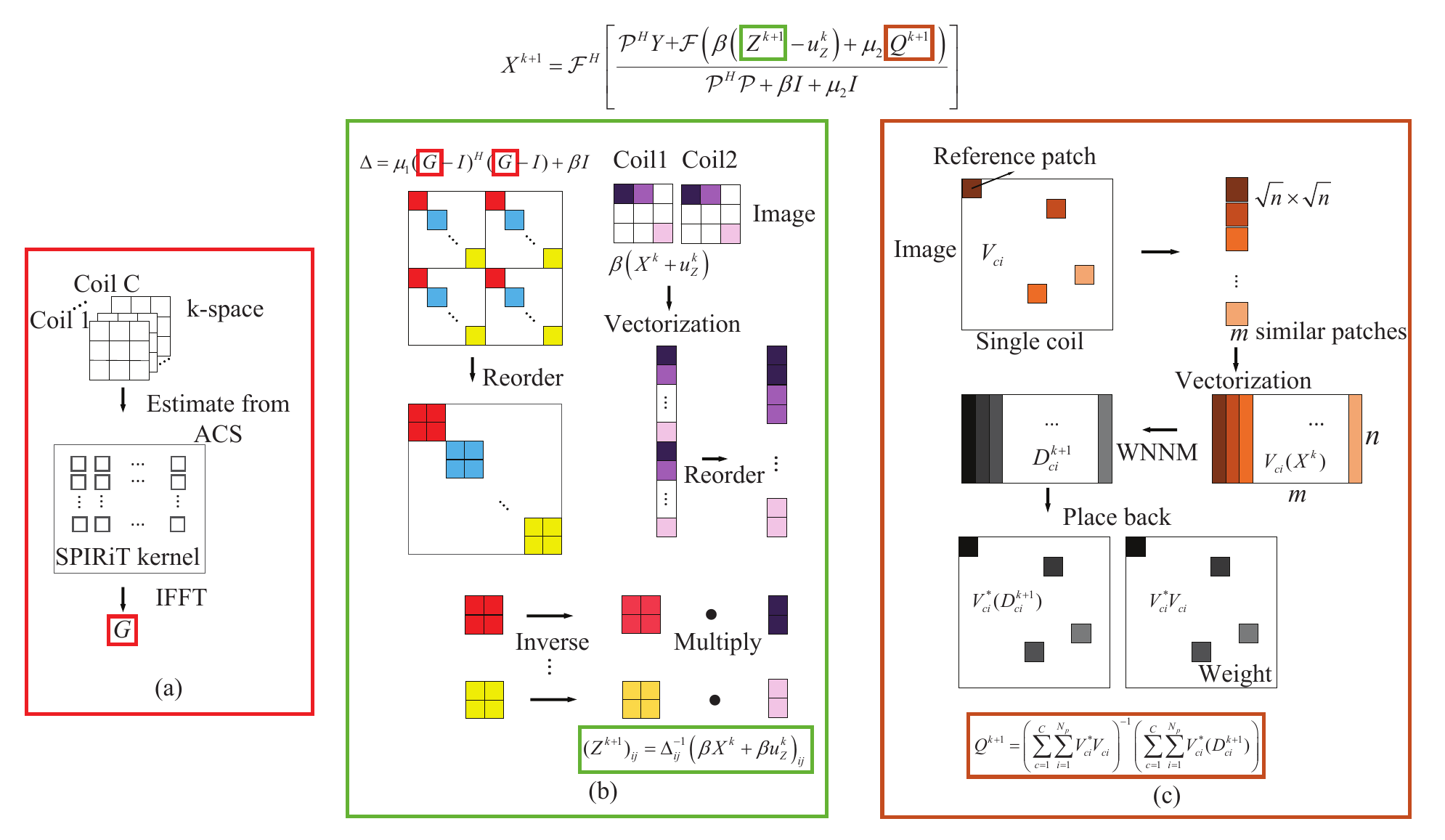}
}
\end{minipage}
}
\vspace{-2.5em} 
\caption{The schematic illustration of NLR-SPIRiT. (a) Calculation of G \cite{Uecker2014}: A series of SPIRiT kernels are estimated through k-space autocalibration signal (ACS). The matrix $G$ is obtained by calculating the IFFT of the matrix composed of SPIRiT kernels. (b) Preprocessing steps for calculating  (take only two coils as an example) \cite{Weller2014}: The matrix $\Delta  = {\mu _1}{(G - I)^H}(G - I) + \beta I$ has a diagonal-block structure, then a simple reordering of the matrix yields a block-diagonal structure. The inverse of the matrix, $\Delta {}^{ - 1}$, is computed by directly inverting each block. (c) Non-local low-rank denoising: Choose  similar patches closest to the reference patch in terms of the Euclidean distance. The selected similar patches are vectorized and grouped to construct a similar patch group matrix ${V_{ci}}(X)$. The low-rank approximation ${D_{ci}}$ is obtained by solving the weighted nuclear norm minimization (WNNM) problem, and then placed back to the original positions.}
\vspace{-0.5em} 
\label{fig0}
}
\end{figure*}

The SPIRiT calibration consistency equation in the image domain is expressed as:
\begin{equation}
X = GX \label{eq2}
\end{equation}
where $G$ is the image domain-based SPIRiT operator, acquired from autocalibration signal (ACS) lines (as shown in Fig. \ref{fig0}(a)) \cite{Uecker2014}.

Then, the image domain-based SPIRiT minimization problem is as follows:
\begin{equation}
X = \arg \;\mathop {\min }\limits_X \frac{1}{2}\left\| {AX - Y} \right\|_2^2 + \frac{\mu }{2}\left\| {(G - I)X} \right\|_2^2 \label{eq3}
\end{equation}

\section{The proposed algorithm}
\label{algorithm}

\subsection{Problem formulation}

Past works have verified that the NSS prior information facilitated sparsity enforcement, resulting in reconstruction quality enhancement. To further improve the PMRI reconstruction quality, we incorporate the NLR regularization term into the SPIRiT model to fully utilize the NSS of PMRI.

Suppose ${X_c} \in \mathbb{C}{^N}$ denotes the $c^{\rm{th}}$ coil image of $X$, and a single coil image is divided into  ${N_p}$ overlapping patches of size  $\sqrt n \times \sqrt n $. For each reference patch, we search for similar patches within a local window (e.g., 40$\times$40) \cite{Dong2014}, and choose $m$ similar patches closest to the reference patch in terms of the Euclidean distance. The selected similar patches are vectorized and grouped to construct a similar patch group matrix  ${V_{ci}}(X) \in \mathbb{C}{^{n \times m}}$, where ${V_{ci}}: \ X \in \mathbb{C}{^{NC}} \mapsto {V_{ci}}(X) \in \mathbb{C} {^{n \times m}}$ is a BM operator. The step of obtaining NSS prior information is depicted in Fig. \ref{fig0}(c). Therefore, the PMRI reconstruction problem based on the NLR regularization term and the SPIRiT model can be formulated as the following problem:
\begin{equation}
\begin{aligned}
X = \arg\mathop {\min }\limits_X \frac{1}{2}\left\| {AX - Y} \right\|_2^2 + \frac{{{\mu _1}}}{2}\left\| {(G - I)X} \right\|_2^2 + \tau \sum\limits_{c = 1}^C {\sum\limits_{i = 1}^{{N_p}} {{\mathop{\rm rank}\nolimits} \left( {{V_{ci}}(X)} \right)} }\label{eq10}
\end{aligned}
\end{equation}
where  ${\mathop{\rm rank}\nolimits} \left( {{V_{ci}}(X)} \right)$ denotes the rank of the matrix  ${V_{ci}}(X)$, ${\mu _1}$  and  $\tau $ are tuning parameters, which are applied to balance the data fidelity, the calibration consistency, and the NLR regularization term.

\subsection{Problem solution}
\label{solution}

Problem (\ref{eq10}) is a large-scale nonconvex optimization problem that is difficult to solve. Authors \cite{Danielyan2012, Yoon2014} have proposed employing an NE formulation. With the application of the NE formulation, we convert the reconstruction (\ref{eq10}) into a two-objective optimization problem:
\begin{equation}
\left\{ {D_{ci}^{k + 1}} \right\} \!=\! \arg\mathop {\min }\limits_{\left\{ {{D_{ci}}} \right\}} \frac{1}{2}\left\| {{V_{ci}}({X^k}) \!-\! {D_{ci}}} \right\|_F^2 \!+\! {\tau} {\mathop{\rm rank}\nolimits} ({D_{ci}})\label{eq11}
\end{equation}
\begin{equation}
\begin{aligned}
{X^{k + 1}} = \arg \mathop {\min }\limits_X \left\| {AX - Y} \right\|_2^2 + {\mu _1}\left\| {\left( {G - I} \right)X} \right\|_2^2 + {\mu _2}\left\| {X - {Q^{k + 1}}} \right\|_2^2\label{eq12}
\end{aligned}
\end{equation}
where  ${\left\|  \cdot  \right\|_F}$ denotes the Frobenius norm, $D_{ci}^{k + 1}$ denotes LR approximation of the patch group matrix  ${V_{ci}}({X^k})$, and $Q^{k + 1}$ is determined as follows:
\begin{equation}
{Q^{k + 1}} \!=\! {\left( {\sum\limits_{c = 1}^C {\sum\limits_{i = 1}^{{N_p}} {V_{ci}^{\rm{*}}{V_{ci}}} } } \! \right) ^{ - 1}} \! \left( {\sum\limits_{c = 1}^C {\sum\limits_{i = 1}^{{N_p}} {V_{ci}^{\rm{*}}(D_{ci}^{k + 1})} } } \! \right) \! \in \! \mathbb{C}{^{NC}}\label{eq13}
\end{equation}
where  $V_{ci}^{\rm{*}}: \ \mathbb{C}{^{n \times m}} \mapsto \mathbb{C} {^{NC}}$, the adjoint operator of  ${V_{ci}}$, places back the denoised patches  $D_{ci}^{k + 1}$ at their original positions (as shown in Fig. \ref{fig0}(c)).  $\sum\nolimits_{c = 1}^C {\sum\nolimits_{i = 1}^{{N_p}} {V_{ci}^{\rm{*}}{V_{ci}}} }  \in \mathbb{C}{^{NC \times NC}}$ is a diagonal matrix with each diagonal element equal to the time of the corresponding pixel belonging to the overlapping patches throughout $\{ {V_{ci}}(X)\} $.

\textbf{1) Minimization with respect to  $\left\{ {{D_{ci}}} \right\}$:} Generally, the rank penalty objective optimization problem of $\left\{ {{D_{ci}}} \right\}$ is a nondeterministic polynomial time (NP)-hard problem. Thus, given that the WNN \cite{Gu2014, Lu2016} may yield a better rank approximation than the NN, we adopt the WNN as a convex surrogate of the rank. The WNN of $\left\{ {{D_{ci}}} \right\}$ can be written as \cite{Gu2014}:

\begin{equation}
{\left\| {{D_{ci}}} \right\|_{w, * }} = \sum\limits_{j = 1}^{\min (n,m)} {{w_j}{\sigma _j}} \left( {{D_{ci}}} \right)\label{eq15}
\end{equation}

With the use of the WNN as a surrogate of the rank, let $2 \tau = {\delta ^2}$, problem (\ref{eq11}) can be rewritten as follows:
\begin{equation}
\{ D_{ci}^{k + 1}\}  = \arg \mathop {\min }\limits_{\{ {D_{ci}}\} } \left\| {{V_{ci}}({X^k}) - {D_{ci}}} \right\|_F^2 + {\delta ^2}{\left\| {{D_{ci}}} \right\|_{w, * }}\label{eq17}
\end{equation}

Problem (\ref{eq17}) is a weighted nuclear norm minimization (WNNM) \cite{Gu2014} problem. Let $U\Sigma {V^H} = {V_{ci}}({X^k})$ be the full singular value decomposition (SVD) of ${V_{ci}}({X^k})$, $\Sigma  = {\mathop{\rm diag}\nolimits} ({\sigma _1}({V_{ci}}({X^k})),...,{\sigma _j}({V_{ci}}({X^k})),...,{\sigma _J}({V_{ci}}({X^k})))$ , ${\sigma _j}\left( {{V_{ci}}({X^k})} \right)$ is the $j^{\rm{th}}$ singular value of ${V_{ci}}({X^k})$, and $J = \min (m,n)$. Hence, the optimal solution to (\ref{eq17}) is  $D_{ci}^{k + 1} = U\Gamma {V^T}$, $\Gamma  = {\mathop{\rm diag}\nolimits} ({\gamma _1},...,{\gamma _j},...,{\gamma _J})$, where ${\gamma _j}$ can be calculated as:

\begin{equation}
{\gamma _j} = {\mathop{\rm soft}\nolimits} \left( {{\sigma _j}\left( {{V_{ci}}({X^k})} \right),{w_j}} \right) \label{eq18}
\end{equation}
where  ${\mathop{\rm soft}\nolimits} (\cdot )$ is the soft threshold operator, ${\mathop{\rm soft}\nolimits} (\sigma ,w) = \max (\sigma  - w,0)$. And the weight  ${w_j}$ can be calculated as:

\begin{equation}
{w_j} = \frac{{b_0\sqrt m }}{{{{\hat \sigma }_j} + \varepsilon }}\label{eq16}
\end{equation}
where $b_0$ is a constant, $\varepsilon  = {10^{ - 16}}$ is to avoid dividing by zero, and then the initial  ${\hat \sigma _j}$ can be initialized by:

\begin{equation}
{\hat \sigma _j} = \sqrt {\max ({\sigma _j}{{({V_{ci}}({X^k}))}^2} - m{\delta ^2},0)} \label{eq16b}
\end{equation}

\floatstyle{ruled}
\newfloat{Algorithm}{!t}{lop}
\begin{Algorithm}
\footnotesize{
\newcommand{\uw}{0.4}
\algsetup{linenosize=\normalsize}
\caption{NLR regularization-based SPIRiT (NLR-SPIRiT) PMRI reconstruction algorithm}
\label{alg1}

\begin{algorithmic}[1]       \vspace{\uw em}
\STATE Input: undersampled k-space data $Y$ of the multicoil image.
\STATE Set ${X^0} = {F^{ - 1}}Y$, ${Z^0} = 0$, $u_Z^0 = 0$, $k=0$, ${\Delta ^{ - 1}} = {\left( {{\mu _1}{{(G - I)}^H}(G - I) + \beta I} \right)^{ - 1}}$                                        \label{alg1:js1} \vspace{\uw em}

\REPEAT                                  \label{alg1:js2} \vspace{\uw em}
\STATE If $\bmod (k,T) = 0$, update the patch grouping ${V_{ci}}$ by BM. \label{alg1:js3} \vspace{\uw em}
\STATE Construct similar patch group matrix ${V_{ci}}({X^k})$. \label{alg1:js4} \vspace{\uw em}

\FOR{ each group ${V_{ci}}({X^k})$ }                    \label{alg1:js5} \vspace{\uw em}
\STATE  Compute the full SVD of  $U\Sigma {V^T}$ of  ${V_{ci}}({X^k})$. \label{alg1:js6} \vspace{\uw em}
\STATE  Update the weights  ${w_j}$ via Eq. (\ref{eq16}). \label{alg1:js7} \vspace{\uw em}
\STATE  Compute $\Gamma  = {\mathop{\rm diag}\nolimits} ({\gamma _1},...,{\gamma _j},...,{\gamma _J})$, where ${\gamma _j}$ was computed via Eq. (\ref{eq18}).  \label{alg1:js8} \vspace{\uw em}
\STATE  Compute  $D_{ci}^{k + 1} = U\Gamma {V^T}$.      \label{alg1:js9} \vspace{\uw em}
\ENDFOR

\STATE Compute  ${Q^{k + 1}}$ via Eq. (\ref{eq13}).     \label{alg1:js10} \vspace{\uw em}
\STATE Compute  ${Z^{k + 1}} = {\Delta ^{ - 1}}\left( {\beta {X^k} + \beta u_Z^k} \right)$. \label{alg1:js11} \vspace{\uw em}
\STATE Compute  ${X^{k + 1}}$ via Eq. (\ref{eq23}).     \label{alg1:js12} \vspace{\uw em}
\STATE Update  $u_Z^{k + 1}$ via Eq. (\ref{eq21}).      \label{alg1:js13} \vspace{\uw em}
\STATE Compute  ${x^{k + 1}}$ via Eq. (\ref{eq23b}).      \label{alg1:js14} \vspace{\uw em}
\STATE $k=k+1$.   \label{alg1:js15} \vspace{\uw em}
\UNTIL $RE < tol$ or $k>K$ \label{alg1:js16} \vspace{\uw em}

\STATE Output the reconstructed image $\hat X$.         \label{alg1:js17} \vspace{\uw em}
\end{algorithmic}
}
\end{Algorithm}

\textbf{2) Image reconstruction:} After calculating  $\{ D_{ci}^{k + 1}\} $, the whole image can be reconstructed by solving problem (\ref{eq12}). By introducing an auxiliary variable  $Z = X$ and corresponding Lagrange multiplier  ${u_Z}$, the LS problem (\ref{eq12}) can be converted to the following subproblems via the ADMM technique:
\begin{equation}
{Z^{k + 1}} \!=\! \arg \mathop {\min }\limits_Z {\mu _1}\left\| {\left( {G \!-\! I} \right)Z} \right\|_2^2 + \beta \!\left\| {Z \!-\! \left( {{X^k} \!+\! u_Z^k} \right)} \right\|_2^2\label{eq19}
\end{equation}
\begin{equation}
\begin{aligned}
{X^{k + 1}} = \arg \mathop {\min }\limits_X \left\| {AX - Y} \right\|_2^2 + \beta \left\| {X - {Z^{k{\rm{ + }}1}} + u_Z^k} \right\|_2^2 + {\mu _2}\left\| {X - {Q^{k + 1}}} \right\|_2^2\label{eq20}
\end{aligned}
\end{equation}
\begin{equation}
u_Z^{k + 1} = u_Z^k + \eta \left( {{X^{k + 1}} - {Z^{k + 1}}} \right) \label{eq21}
\end{equation}

The solution to subproblem (\ref{eq19}) with respect to  $Z$ is given by:
\begin{equation}
{Z^{k + 1}} = {\left[ {{\mu _1}{{(G \!-\! I)}^H}(G \!-\! I) + \beta I} \right]^{ - 1}} \!\! \left( {\beta {X^k} + \beta u_Z^k} \right) \label{eq22}
\end{equation}

Let  $\Delta  = {\mu _1}{(G - I)^H}(G - I) + \beta I$, and  ${\Delta ^{ - 1}}$ can thus be obtained via the direct inversion of each  $C \times C$ block in matrix  $\Delta $, as shown in Fig. \ref{fig0}(b) \cite{Weller2014}. Then, we have the update of  $Z$ as  ${Z^{k + 1}} = {\Delta ^{ - 1}}\left( {\beta {X^k} + \beta u_Z^k} \right)$.
 ${{\cal P}^H}{\cal P}$ is a diagonal matrix, and subproblem (\ref{eq20}) with respect to  $X$ yields the following closed-form solution (please refer to equations (\ref{eq32})-(\ref{eq35}) of appendix for the similar derivation of $X$):
\begin{equation}
{X^{k + 1}} = {{\cal F}^H} \! \left[ {\frac{{{{\cal P}^H}Y{\rm{ + }}{\cal F}\left( {\beta \left( {{Z^{k + 1}} \!-\! u_Z^k} \right) + {\mu _2}{Q^{k + 1}}} \right)}}{{{{\cal P}^H}{\cal P} + \beta I + {\mu _2}I}}} \right] \label{eq23}
\end{equation}

Since subproblems (\ref{eq11}), (\ref{eq19}), and (\ref{eq20}) are efficiently solved, the  C-coil image $X$ is obtained, and then combined into a single magnitude image $x$ by using the square root of sum of squares (SOS):
\begin{equation}
x = {\rm{SOS}}(X)= \sqrt {\sum\limits_{c = 1}^C {{{\left| {{X_c}} \right|}^2}} } \label{eq23b}
\end{equation}

Finally, we obtain the NLR regularization-based SPIRiT (NLR-SPIRiT) PMRI reconstruction algorithm, as expressed in Algorithm \ref{alg1}. And the schematic illustration of NLR-SPIRiT is described in Fig. \ref{fig0}. Algorithm \ref{alg1} is terminated when the relative error (RE) $RE = {{{{\left\| {{x^{k + 1}} - {x^k}} \right\|}_2}} \mathord{\left/
 {\vphantom {{{{\left\| {{x^{k + 1}} - {x^k}} \right\|}_2}} {{{\left\| {{x^k}} \right\|}_2}}}} \right.
 \kern-\nulldelimiterspace} {{{\left\| {{x^k}} \right\|}_2}}}$ falls below the tolerance $tol$. In Algorithm \ref{alg1}, the BM step is performed every $T$ ($T=3$) iterations to reduce the computational complexity, as shown in step \ref{alg1:js3}.

The optimization problem (\ref{eq10}) can also be solved by using the variable splitting and ADMM techniques \cite{Afonso2010}. The corresponding algorithm is called ADMM-NLR-SPIRiT. Please refer to the appendix for details.

\section{Experimental results}
\label{result}

\subsection{Experimental setup}
\label{result:setup}

In our experiments, we compared the proposed NLR-SPIRiT model to state-of-the-art algorithms to solve the PMRI reconstruction problems such as: JTV-SPIRiT \cite{Weller2014}, STDLR-SPIRiT \cite{Zhang2020}, pFISTA-SPIRiT \cite{Zhang2021}, and NLR-SPIRiT-baseline (a variant of the NLR-SPIRiT model with the standard NN). All the considered algorithms were implemented in MATLAB. The source code of NLR-SPIRiT and data can be downloaded from the following website: \url{https://drive.google.com/file/d/1Pkw__GA9nzOqYFuk4gWkGa3rztHaEy} \\ \url{J4/view?usp=sharing}.

To validate the performances of all the considered algorithms, we conduct experiments on freely available fully sampled in vivo human datasets. The first set of data include 47 brain datasets \cite{Souza2018} (where dataset 1 is shown in Fig. \ref{fig3}\subref{fig3a} and other six datasets are shown in Fig. \ref{fig10}). All data are acquired on a clinical MR scanner (Discovery MR750) using a 12-channel head-neck-spine coil and a 3D T1-weighted gradient echo sequence (matrix size = $256\times218\times170$ or $256\times218\times180$, TR/TE/TI = 6.3/2.6/650 ms or TR/TE/TI = 7.4/3.1/400 ms, slice thickness = 1 mm). We chose a single slice of each multislice dataset in our experiment. The second fully sampled knee dataset \cite{Epperson2013} (dataset 2, as shown in Fig. \ref{fig4}\subref{fig4a}) was acquired on a GE scanner using an 8-channel HD knee coil and a 3D FSE CUBE sequence (matrix size = $320\times320\times256$, TR/TE = 1550/25 ms, FOV = $160\times160$ mm, slice thickness = 0.6 mm). We chose a single slice of each multislice dataset in our experiment. The third fully sampled brain dataset \cite{Lin2008} (dataset 3, as shown in Fig. \ref{fig5}\subref{fig5a}) was acquired on a 3T SIEMENS Trio system and an 8-channel head array coil (matrix size = $256\times256$, TR/TE = 2530/3.45 ms, slice thickness = 1.33 mm, FOV = $256\times256$ mm).

To validate the considered algorithms, fully sampled datasets were subjected to retrospective undersampling with the above different undersampling patterns in our experiments, such as the 2DPU patterns (including $24\times24$ ACS lines), the 1DUU patterns and the 1DGU patterns (including 20 ACS lines) with different acceleration factors (AFs). All experiments were simulated on a workstation machine equipped with an Intel Core i9-10900X @ 3.7 GHz processor, 64 GB of RAM memory and a Windows 10 operating system (64 bit).

Three commonly adopted objective metrics were employed to evaluate the quality of the reconstructed images, namely, the signal-to-noise ratio (SNR) \cite{Chartrand2009}, high-frequency error norm (HFEN) \cite{ Ravishankar2011}, and structural similarity index measure (SSIM) \cite{Wang2004}. It should be noted that the calculation of all metrics is limited to the region of interest (ROI). High SNR and SSIM values or low HFEN values indicate more accurate reconstruction. In regard to the reference image  $x$ and the reconstructed image $\hat x$, the SNR is defined as:
\begin{equation}
SNR = 10{\log _{10}}\left( {\frac{{Var}}{{MSE}}} \right)\label{eq33}
\end{equation}
where  $MSE$ denotes the mean square error between $x$  and $\hat x$, and $Var$ is the variance in $x$.

The HFEN is expressed as:
\begin{equation}
HFEN = \frac{{{{\left\| {filter(\hat x) - filter(x)} \right\|}_2}}}{{{{\left\| {filter(x)} \right\|}_2}}}\label{eq34}
\end{equation}
where  $filter(.)$ is a Laplacian Gaussian filter operator used to determine the image edges.

The SSIM is calculated as:
\begin{equation}
SSIM = \frac{{\left( {2{u_x}{u_{\hat x}} + {c_1}} \right)\left( {2{\sigma _{x\hat x}} + {c_2}} \right)}}{{\left( {{u_x}^2 + {u_{\hat x}}^2 + {c_1}} \right)\left( {{\sigma _x}^2 + {\sigma _{\hat x}}^2 + {c_1}} \right)}} \label{eq35}
\end{equation}
where ${u_x}$ and ${u_{\hat x}}$ are the means of $x$ and $\hat x$, respectively, ${\sigma _x}^2$ and ${\sigma _{\hat x}}^2$ are the variances of $x$ and $\hat x$, respectively, ${\sigma _{x\hat x}}$ represents the covariance of $x$ and $\hat x$, ${c_1}$ and  ${c_2}$ are constant, where ${c_1}=0.01$ and ${c_2}=0.03$.

\subsection{Parameter settings}

A kernel size of $5\times 5$ was adopted in the SPIRiT-based algorithms in the following experiments. The parameters of JTV-SPIRiT, STDLR-SPIRiT and pFISTA-SPIRiT were manually tuned for SNR optimal. In regard to NLR-SPIRiT-baseline and NLR-SPIRiT, $6\times 6$ $(n = 36)$ image patches were used, and the total number of similar patches was 43 $(m=43)$. A reference image patch was extracted every 5 pixels along the horizontal and vertical directions to reduce the computational complexity. The main parameters were defined as follows: ${\mu _1} = {\mu _2} = 1$, $b_0=0.4$, and $\eta=\sqrt 2$. These settings remained fixed in all experiments.

In addition, to obtain a better parallel image reconstruction performance, the parameter  $\delta $  and $\beta $ were adjusted to optimize the SNR and HFEN values. We run the proposed algorithm on the same undersampled data considering the ranges of  ${\delta} \in \left[ {1,6} \right]$ and  $\beta  \in [0.1,1]$.

We chose ${\delta}$  and $\beta $  based on the highest SNR and smallest HFEN values. For example, at ${\delta} = 3$ and $\beta  = 0.3$, these two optimal indicators (SNR and HFEN) of dataset 1 were roughly optimized.

\subsection{Convergence analysis}
\begin{figure}[!t]
\footnotesize{
\centering{ %
\newcommand{\uwidth}{1.7} 
\newcommand{\uhoriz}{-0.6}  
\begin{minipage}[b]{ 0.976 \textwidth} %
\centering{
\subfloat[]{\includegraphics[width= \uwidth in]{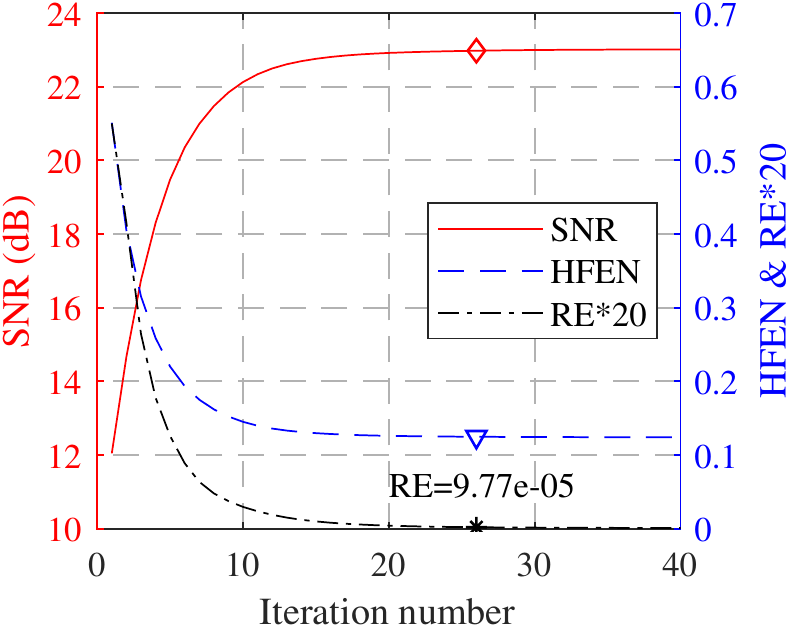}
\label{fig2a}}
\hspace{\uhoriz em}
\subfloat[]{\includegraphics[width= \uwidth in]{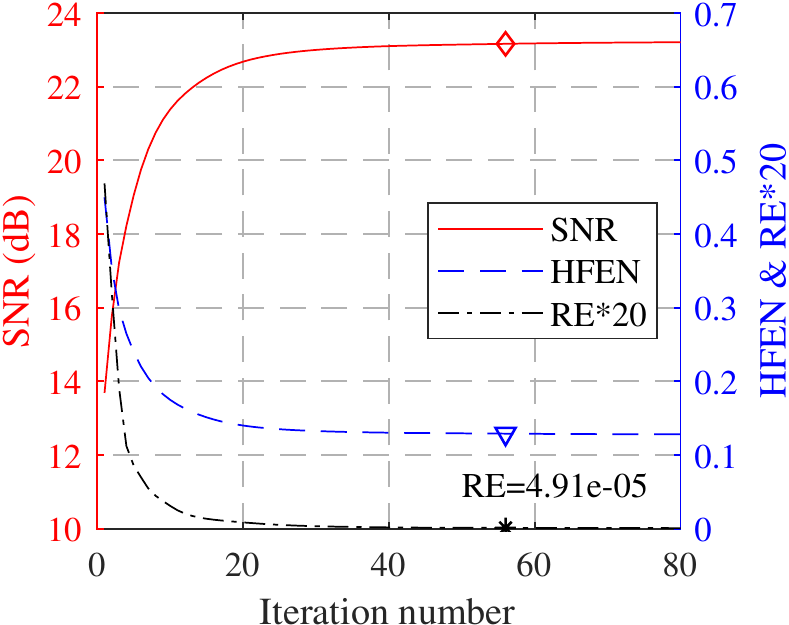}
\label{fig2b}}
\hspace{\uhoriz em}
\subfloat[]{\includegraphics[width= \uwidth in]{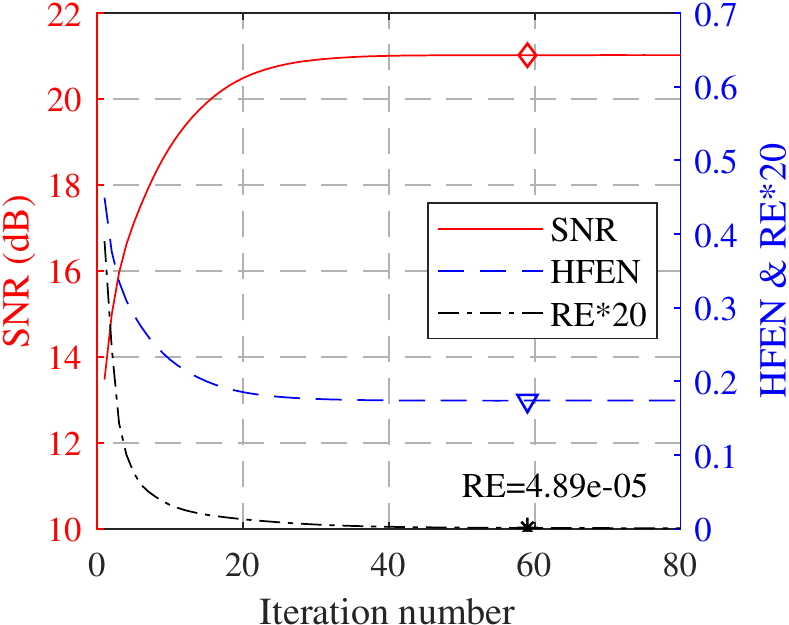}
\label{fig2c}}
}
\end{minipage}
}
\vspace{-1em} 
\caption{SNR, HFEN and RE versus the iteration number when reconstructing the dataset 1. (a), (b), and (c) are based on the 2DPU pattern ($AF=5$), the 1DUU pattern ($AF=3$), and the 1DGU pattern ($AF=3$), respectively.}
\vspace{-0.5em} 
\label{fig2}
}
\end{figure}

To reflect the convergence of the proposed NLR-SPIRiT algorithm, Fig. \ref{fig2} shows SNR, HFEN, and RE versus the iteration number during the reconstruction of dataset 1 based on the 2DPU pattern $(AF=5)$, the 1DUU pattern $(AF=3)$, and the 1DGU pattern $(AF=3)$. As shown in Fig. \ref{fig2}(a), NLR-SPIRiT approximately reaches the maximum SNR and the minimum HFEN when RE falls below $1e-4$. In Fig. \ref{fig2}(b) and Fig. \ref{fig2}(c), the approximately maximum SNR and the minimum HFEN of NLR-SPIRiT occur when RE falls below $5e-5$. At this time, NLR-SPIRiT converges. Moreover, almost all the tested data can achieve the maximum SNR at $K=30$  and $K=80$ based on 2D and 1D undersampling pattern, respectively. Hence, $K=30$ , $tol=1e-4$  and $K=80$ , $tol=5e-5$ are selected as the maximum iteration and the stop tolerance for the 2DPU patterns and 1D undersampling patterns, respectively.

\begin{figure*}[!t]
\footnotesize{
\centering{ %

\newcommand{\uwidth}{0.85}   
\newcommand{\uhoriz}{-0.5}  

\begin{minipage}[b]{ 0.976 \textwidth} %
\centering{
\subfloat[]{\includegraphics[width= \uwidth in]{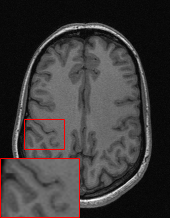}
\label{fig3a}}
\hspace{\uhoriz em}
\subfloat[]{\includegraphics[width= \uwidth in]{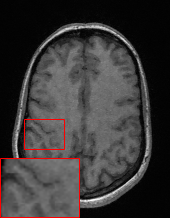}
\label{fig3b}}
\hspace{\uhoriz em}
\subfloat[]{\includegraphics[width= \uwidth in]{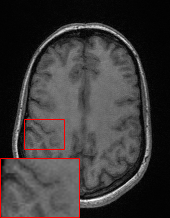}
\label{fig3c}}
\hspace{\uhoriz em}
\subfloat[]{\includegraphics[width= \uwidth in]{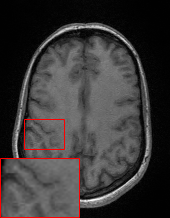}
\label{fig3d}}
\hspace{\uhoriz em}
\subfloat[]{\includegraphics[width= \uwidth in]{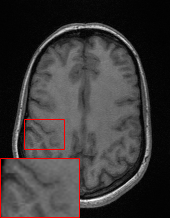}
\label{fig3e}}
\hspace{\uhoriz em}
\subfloat[]{\includegraphics[width= \uwidth in]{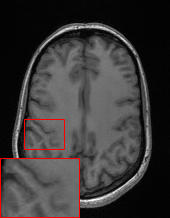}
\label{fig3f}}
\\
\vspace{-0.5 em}
\subfloat[]{\includegraphics[width= \uwidth in]{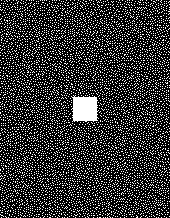}
\label{fig3g}}
\hspace{\uhoriz em}
\subfloat[]{\includegraphics[width= \uwidth in]{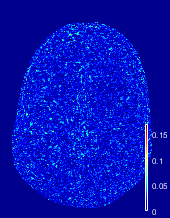}
\label{fig3h}}
\hspace{\uhoriz em}
\subfloat[]{\includegraphics[width= \uwidth in]{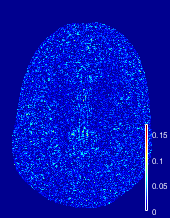}
\label{fig3i}}
\hspace{\uhoriz em}
\subfloat[]{\includegraphics[width= \uwidth in]{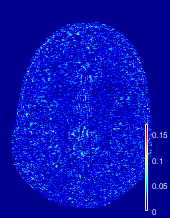}
\label{fig3j}}
\hspace{\uhoriz em}
\subfloat[]{\includegraphics[width= \uwidth in]{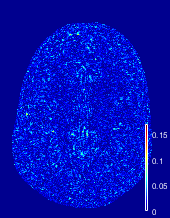}
\label{fig3k}}
\hspace{\uhoriz em}
\subfloat[]{\includegraphics[width= \uwidth in]{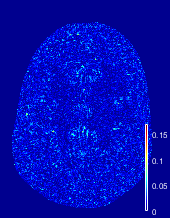}
\label{fig3l}}
}
\end{minipage}
}
\vspace{-0.2em} 
\caption{Reconstructed results of dataset 1 based on the 2DPU pattern $(AF=5)$. (a) Reference image; (b) reconstructed image via JTV-SPIRiT \cite{Weller2014}; (c) reconstructed image via LPJTV-ESPIRiT \cite{Duan2019}; (d) reconstructed image via STDLR-SPIRiT \cite{Zhang2020}; (e) reconstructed image via the NLR-SPIRiT-baseline; (f) reconstructed image by NLR-SPIRiT; (g) undersampling pattern; (h), (i), (j), (k), and (l) show the error maps of (b), (c), (d), (e), and (f), respectively.}
\vspace{-0.5em} 
\label{fig3}
}
\end{figure*}

\begin{figure*}[!t]
\footnotesize{
\centering{ %

\newcommand{\uwidth}{0.85}   
\newcommand{\uhoriz}{-0.5}  

\begin{minipage}[b]{ 0.976 \textwidth} %
\centering{
\subfloat[]{\includegraphics[width= \uwidth in]{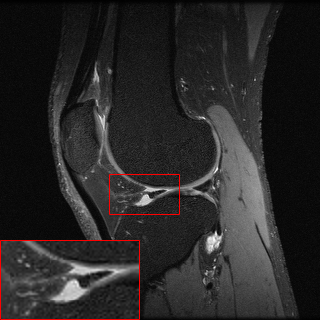}
\label{fig4a}}
\hspace{\uhoriz em}
\subfloat[]{\includegraphics[width= \uwidth in]{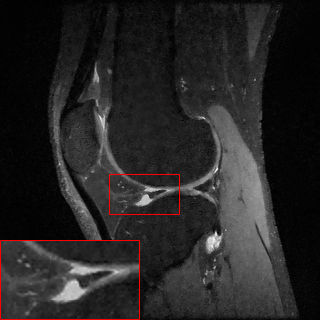}
\label{fig4b}}
\hspace{\uhoriz em}
\subfloat[]{\includegraphics[width= \uwidth in]{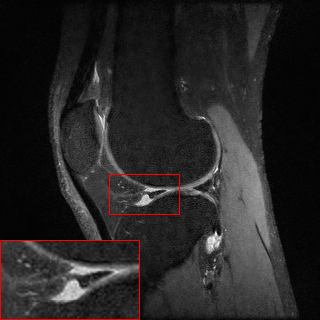}
\label{fig4c}}
\hspace{\uhoriz em}
\subfloat[]{\includegraphics[width= \uwidth in]{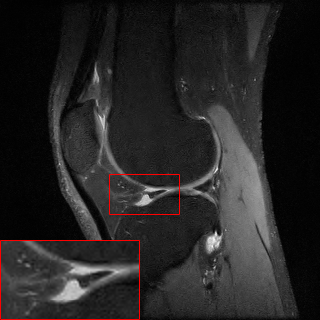}
\label{fig4d}}
\hspace{\uhoriz em}
\subfloat[]{\includegraphics[width= \uwidth in]{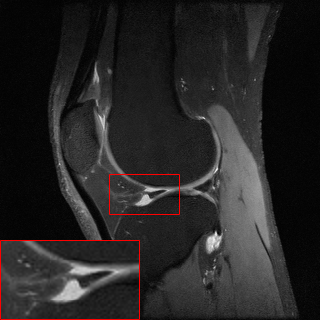}
\label{fig4e}}
\hspace{\uhoriz em}
\subfloat[]{\includegraphics[width= \uwidth in]{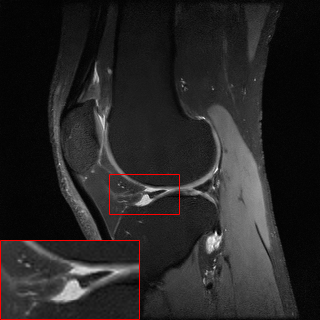}
\label{fig4f}}
\\
\vspace{-0.5 em}
\subfloat[]{\includegraphics[width= \uwidth in]{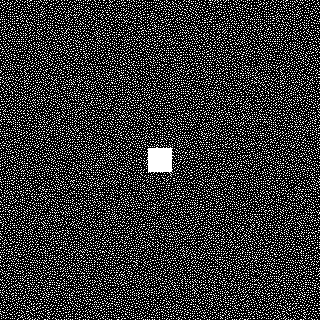}
\label{fig4g}}
\hspace{\uhoriz em}
\subfloat[]{\includegraphics[width= \uwidth in]{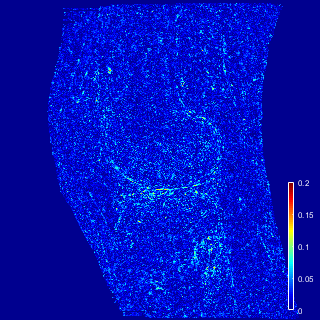}
\label{fig4h}}
\hspace{\uhoriz em}
\subfloat[]{\includegraphics[width= \uwidth in]{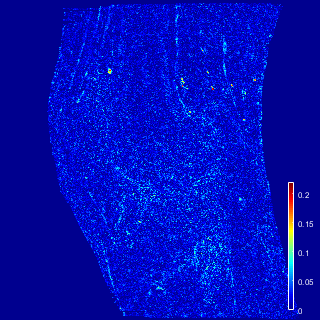}
\label{fig4i}}
\hspace{\uhoriz em}
\subfloat[]{\includegraphics[width= \uwidth in]{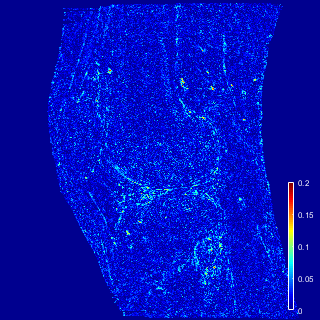}
\label{fig4j}}
\hspace{\uhoriz em}
\subfloat[]{\includegraphics[width= \uwidth in]{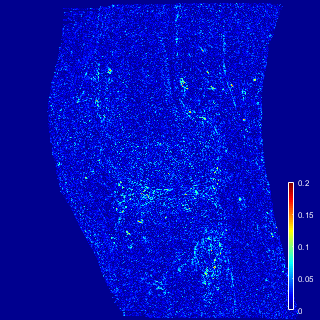}
\label{fig4k}}
\hspace{\uhoriz em}
\subfloat[]{\includegraphics[width= \uwidth in]{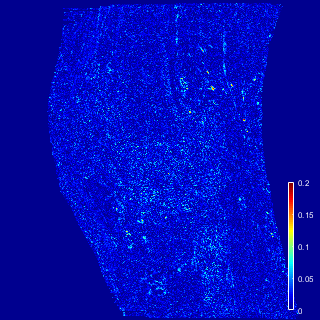}
\label{fig4l}}
}
\end{minipage}
}
\vspace{-0.2em} 
\caption{Reconstructed results of dataset 2 based on the 2DPU pattern $(AF=5)$. (a) Reference image; (b) reconstructed image via JTV-SPIRiT \cite{Weller2014}; (c) reconstructed image via LPJTV-ESPIRiT \cite{Duan2019}; (d) reconstructed image via STDLR-SPIRiT \cite{Zhang2020}; (e) reconstructed image via the NLR-SPIRiT-baseline; (f) reconstructed image by NLR-SPIRiT; (g) undersampling pattern; (h), (i), (j), (k), and (l) show the error maps of (b), (c), (d), (e), and (f), respectively.}
\vspace{-0.5em} 
\label{fig4}
}
\end{figure*}

\begin{figure*}[!t]
\footnotesize{
\centering{ %

\newcommand{\uwidth}{0.85}   
\newcommand{\uhoriz}{-0.5}  

\begin{minipage}[b]{ 0.976 \textwidth} %
\centering{
\subfloat[]{\includegraphics[width= \uwidth in]{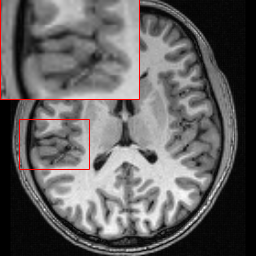}
\label{fig5a}}
\hspace{\uhoriz em}
\subfloat[]{\includegraphics[width= \uwidth in]{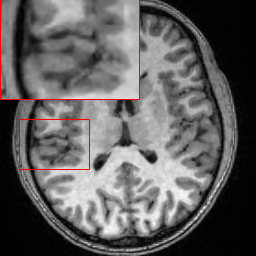}
\label{fig5b}}
\hspace{\uhoriz em}
\subfloat[]{\includegraphics[width= \uwidth in]{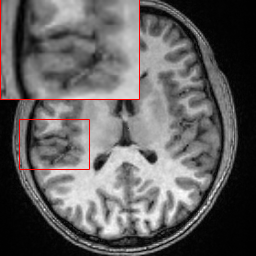}
\label{fig5c}}
\hspace{\uhoriz em}
\subfloat[]{\includegraphics[width= \uwidth in]{data4_enlarge_STDLR_SPIRiT_R5}
\label{fig5d}}
\hspace{\uhoriz em}
\subfloat[]{\includegraphics[width= \uwidth in]{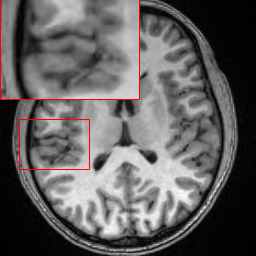}
\label{fig5e}}
\hspace{\uhoriz em}
\subfloat[]{\includegraphics[width= \uwidth in]{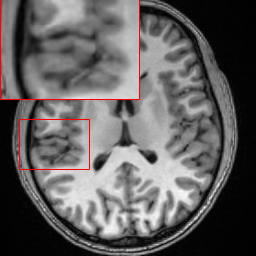}
\label{fig5f}}
\\
\vspace{-0.5 em}
\subfloat[]{\includegraphics[width= \uwidth in]{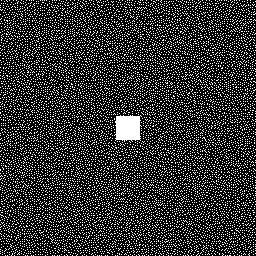}
\label{fig5g}}
\hspace{\uhoriz em}
\subfloat[]{\includegraphics[width= \uwidth in]{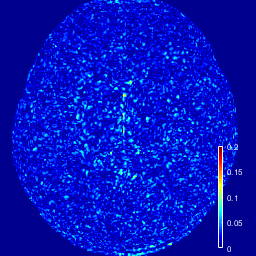}
\label{fig5h}}
\hspace{\uhoriz em}
\subfloat[]{\includegraphics[width= \uwidth in]{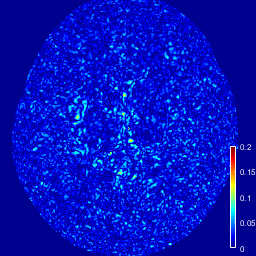}
\label{fig5i}}
\hspace{\uhoriz em}
\subfloat[]{\includegraphics[width= \uwidth in]{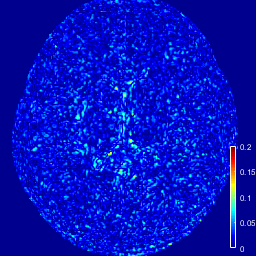}
\label{fig5j}}
\hspace{\uhoriz em}
\subfloat[]{\includegraphics[width= \uwidth in]{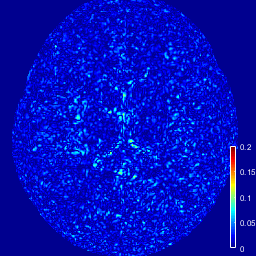}
\label{fig5k}}
\hspace{\uhoriz em}
\subfloat[]{\includegraphics[width= \uwidth in]{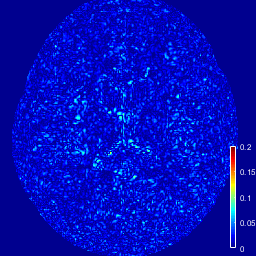}
\label{fig5l}}
}
\end{minipage}
}
\vspace{-0.2em} 
\caption{Reconstructed results of dataset 3 based on the 2DPU pattern $(AF=5)$. (a) Reference image; (b) reconstructed image via JTV-SPIRiT \cite{Weller2014}; (c) reconstructed image via LPJTV-ESPIRiT \cite{Duan2019}; (d) reconstructed image via STDLR-SPIRiT \cite{Zhang2020}; (e) reconstructed image via the NLR-SPIRiT-baseline; (f) reconstructed image by NLR-SPIRiT; (g) undersampling pattern; (h), (i), (j), (k), and (l) show the error maps of (b), (c), (d), (e), and (f), respectively.}
\vspace{-0.5em} 
\label{fig5}
}
\end{figure*}

\subsection{Comparison to previous works}

In the following experiments, JTV-SPIRiT \cite{Weller2014}, STDLR-SPIRiT \cite{Zhang2020}, pFISTA-SPIRiT \cite{Zhang2021}, NLR-SPIRiT-baseline, and NLR-SPIRiT were compared.

\subsubsection{Comparison results of 2DPU patterns}

To facilitate the assessment of the image quality (commonly a subjective parameter), Figs. \ref{fig3}-\ref{fig5} show the images reconstructed via all considered algorithms based on the 2DPU patterns $(AF=5)$, and their corresponding error maps. As shown in Fig. \ref{fig3}, the reconstructed image reconstructed via JTV-SPIRiT obvious artifacts. STDLR-SPIRiT, pFISTA-SPIRiT and NLR-SPIRiT-baseline mitigate these artifacts to a certain extent. The proposed NLR-SPIRiT algorithm effectively removes image artifacts and preserves the details. Hence, NLR-SPIRiT achieves the best visual quality among all competing algorithms. Fig. \ref{fig4} shows that the reconstructed images via JTV-SPIRiT and pFISTA-SPIRiT exhibit apparent artifacts. STDLR-SPIRiT and NLR-SPIRiT-baseline mitigate artifacts to a certain extent. The proposed NLR-SPIRiT algorithm further removes these artifacts and succeeds in reconstructing certain details accurately, such as knee joint, whereas artifact areas remain in images more obviously reconstructed via other considered algorithms. NLR-SPIRiT produces the visually best reconstructed image based on visual comparison. Fig. \ref{fig5} shows that the reconstructed images via JTV-SPIRiT, STDLR-SPIRiT, pFISTA-SPIRiT, and NLR-SPIRiT-baseline exhibit comparable artifacts. NLR-SPIRiT obtain fewest errors and produces the visually best reconstructed image based on visual comparison. In summary, STDLR-SPIRiT and pFISTA-SPIRiT obtain better visual reconstructed performance than JTV-SPIRiT, NLR-SPIRiT-baseline can improve the visual performance of image reconstruction to a certain extent, and NLR-SPIRiT model achieves the best visual performance.

\begin{table}[!h]
\centering
\footnotesize{
\renewcommand{\arraystretch}{1}  
\addtolength{\tabcolsep}{-3pt}    
\caption{Comparison of the three metrics of the reconstructed images via the competing algorithms based on 2DPU patterns $(AF=3-7)$ for dataset 1.}
    \begin{tabular}{ccccccc}
    \hline
    \hline
    \multirow{2}[4]{*}{Algorithms} & \multirow{2}[4]{*}{Metrics} & \multicolumn{5}{c}{Acceleration Factor (AF)} \\
    \cline{3-7} &  & 3 & 4 & 5 & 6 & 7  \\
    \hline
    \multirow{3}[2]{*}{JTV-SPIRiT \cite{Weller2014}} & SNR  & 24.35 & 22.85 & 21.55 & 20.38 & 19.56 \\
          & HFEN & 0.0938 & 0.1197 & 0.1505 & 0.1816 & 0.2048 \\
          & SSIM & 0.9679 & 0.9585 & 0.9483 & 0.9373 & 0.9277 \\
    \hline
    \multirow{3}[2]{*}{STDLR-SPIRiT \cite{Zhang2020}} & SNR  & 24.54 & 22.91 & 21.45 & 20.22 & 19.34 \\
          & HFEN & 0.0898 & 0.1141 & 0.1438 & 0.1715 & 0.1943 \\
          & SSIM & 0.9689 & 0.9577 & 0.9452 & 0.9328 & 0.9220 \\
    \hline
    \multirow{3}[2]{*}{pFISTA-SPIRiT \cite{Zhang2021}} & SNR  & 24.48 & 22.91 & 21.63 & 20.58 & 19.76 \\
          & HFEN & 0.0916 & 0.1193 & 0.146 & 0.1747 & 0.1988 \\
          & SSIM & 0.9706 & 0.9611 & 0.95  & 0.9399 & 0.9311 \\
    \hline
    \multirow{3}[2]{*}{NLR-SPIRiT-baseline} & SNR  & 24.79 & 23.37 & 22.16 & 21.16 & 20.36 \\
          & HFEN & 0.0898 & 0.1122 & 0.1381 & 0.1605 & 0.1804 \\
          & SSIM & 0.9712 & 0.9626 & 0.9532 & 0.9436 & 0.9344 \\
    \hline
    \multirow{3}[2]{*}{NLR-SPIRiT} & SNR  & \textbf{25.37} & \textbf{24.16} & \textbf{22.98} & \textbf{21.90} & \textbf{21.05} \\
          & HFEN & \textbf{0.0810} & \textbf{0.0995} & \textbf{0.1247} & \textbf{0.1509} & \textbf{0.1744} \\
          & SSIM & \textbf{0.9749} & \textbf{0.9690} & \textbf{0.961} & \textbf{0.9516} & \textbf{0.9429} \\
    \hline
	\hline
    \end{tabular}%
  \label{tab2}%
  \vspace{-1em}
  }
\end{table}%

\begin{table}[!h]
\centering
\footnotesize{
\renewcommand{\arraystretch}{1}  
\addtolength{\tabcolsep}{-3pt}    
\caption{Comparison of the three metrics of the reconstructed images via the competing algorithms based on 2DPU patterns $(AF=3-7)$ for dataset 2.}
    \begin{tabular}{ccccccc}
    \hline
	\hline
    \multirow{2}[4]{*}{Algorithms} & \multirow{2}[4]{*}{Metrics} & \multicolumn{5}{c}{Acceleration Factor (AF)} \\
    \cline{3-7} &  & 3 & 4 & 5 & 6 & 7  \\
    \hline
    \multirow{3}[2]{*}{JTV-SPIRiT \cite{Weller2014}} & SNR  & 18.75 & 17.53 & 16.85 & 16.23 & 15.76 \\
          & HFEN & 0.2507 & 0.3022 & 0.3272 & 0.3568 & 0.3855 \\
          & SSIM & 0.9161 & 0.8953 & 0.8831 & 0.8739 & 0.8638 \\
    \hline
    \multirow{3}[2]{*}{STDLR-SPIRiT \cite{Zhang2020}} & SNR  & 19.34 & 17.83 & 17.02 & 16.07 & 15.24 \\
          & HFEN & 0.2245 & 0.2810 & 0.3084 & 0.3483 & 0.3886 \\
          & SSIM & 0.9262 & 0.9026 & 0.8878 & 0.8714 & 0.8524 \\
    \hline
    \multirow{3}[2]{*}{pFISTA-SPIRiT \cite{Zhang2021}} & SNR  & 19.39 & 17.95 & 17.11 & 16.57 & 16.05 \\
          & HFEN & 0.2069 & 0.2654 & 0.3   & 0.3244 & 0.3537 \\
          & SSIM & 0.929 & 0.9069 & 0.892 & 0.8801 & 0.8702 \\
    \hline
    \multirow{3}[2]{*}{NLR-SPIRiT-baseline} & SNR  & 19.13 & 18.42 & 17.75 & 17.20  & 16.67 \\
          & HFEN & 0.2146 & 0.2427 & 0.2655 & 0.2909 & 0.3143 \\
          & SSIM & 0.9201 & 0.9094 & 0.8984 & 0.8889 & 0.8781 \\
    \hline
       \multirow{3}[2]{*}{NLR-SPIRiT} & SNR  & \textbf{20.24} & \textbf{19.08} & \textbf{18.33} & \textbf{17.75} & \textbf{17.19} \\
          & HFEN & \textbf{0.1860} & \textbf{0.2184} & \textbf{0.2421} & \textbf{0.2667} & \textbf{0.2921} \\
          & SSIM & \textbf{0.9352} & \textbf{0.9172} & \textbf{0.9049} & \textbf{0.8947} & \textbf{0.8852} \\
    \hline
	\hline
    \end{tabular}%
  \label{tab3}%
  \vspace{-1em}
  }
\end{table}%

\begin{table}[!h]
\centering
\footnotesize{
\renewcommand{\arraystretch}{1}  
\addtolength{\tabcolsep}{-3pt}    
\caption{Comparison of the three metrics of the reconstructed images via the competing algorithms based on 2DPU patterns $(AF=3-7)$ for dataset 3.}
    \begin{tabular}{ccccccc}
    \hline
	\hline
    \multirow{2}[4]{*}{Algorithms} & \multirow{2}[4]{*}{Metrics} & \multicolumn{5}{c}{Acceleration Factor (AF)} \\
    \cline{3-7} &  & 3 & 4 & 5 & 6 & 7  \\
    \hline
    \multirow{3}[1]{*}{JTV-SPIRiT \cite{Weller2014}} & SNR  & 27.36 & 24.73 & 22.65 & 20.55 & 19.22 \\
          & HFEN & 0.1362 & 0.1836 & 0.2322 & 0.2994 & 0.3382 \\
          & SSIM & 0.9756 & 0.9619 & 0.9463 & 0.9273 & 0.9112 \\
    \hline
    \multirow{3}[2]{*}{STDLR-SPIRiT \cite{Zhang2020}} & SNR  & 27.53 & 25.11 & 22.99 & 20.94 & 19.48 \\
          & HFEN & 0.1333 & 0.1765 & 0.2245 & 0.2882 & 0.3316 \\
          & SSIM & 0.9807 & 0.9702 & 0.9576 & 0.9409 & 0.9250 \\
    \hline
    \multirow{3}[2]{*}{pFISTA-SPIRiT \cite{Zhang2021}} & SNR  & 27.53 & 25.11 & 22.95 & 20.83 & 19.44 \\
          & HFEN & 0.1327 & 0.1776 & 0.2262 & 0.2914 & 0.3343 \\
          & SSIM & 0.9817 & 0.971 & 0.9583 & 0.9413 & 0.9264 \\
    \hline
    \multirow{3}[2]{*}{NLR-SPIRiT-baseline} & SNR  & 27.74 & 25.36 & 23.80  & 21.80  & 20.45 \\
          & HFEN & 0.1294 & 0.1685 & 0.2022 & 0.2586 & 0.2948 \\
          & SSIM & 0.9819 & 0.9727 & 0.9644 & 0.9497 & 0.9381 \\
    \hline
    \multirow{3}[2]{*}{NLR-SPIRiT} & SNR  & \textbf{28.11} & \textbf{26.11} & \textbf{24.31} & \textbf{22.27} & \textbf{20.73} \\
          & HFEN & \textbf{0.1223} & \textbf{0.1544} & \textbf{0.1909} & \textbf{0.2446} & \textbf{0.2849} \\
          & SSIM & \textbf{0.9836} & \textbf{0.9761} & \textbf{0.9673} & \textbf{0.9529} & \textbf{0.9406} \\
    \hline
	\hline
    \end{tabular}%
  \label{tab4}%
  \vspace{-1em}
  }
\end{table}%

To quantitatively evaluate the reconstruction performance of all considered algorithms based on 2DPU patterns with different AF values, comparison results of the SNR, HFEN, and SSIM values of the reconstructed images via the considered algorithms are summarized in Tables \ref{tab2}-\ref{tab4}. The best metric in each case is marked in bold.

As indicated in Tables \ref{tab2}-\ref{tab4}, one can see that the proposed NLR-SPIRiT method produces the best metrics for all datasets and AFs. For dataset 1, NLR-SPIRiT achieves SNR improvements of 1.34 dB, 1.39 dB, 1.21 dB, and 0.71 dB on average over JTV-SPIRiT, STDLR-SPIRiT, pFISTA-SPIRiT and NLR-SPIRiT-baseline, respectively. In regard to dataset 2, NLR-SPIRiT achieves SNR improvements of 1.49 dB, 1.41 dB, 1.10 dB, and 0.68 dB, respectively, on average. Regarding dataset 3, NLR-SPIRiT achieves SNR improvements of 1.40 dB, 1.09 dB, 1.04 dB, and 0.48 dB, respectively, on average.

\subsubsection{Comparison results of 1D undersampling patterns}
\begin{figure*}[!t]
\footnotesize{
\centering{ %

\newcommand{\uwidth}{0.85}   
\newcommand{\uhoriz}{-0.5}  

\begin{minipage}[b]{ 0.976 \textwidth} %
\centering{
\subfloat[]{\includegraphics[width= \uwidth in]{data2_enlarge_I1_R5}
\label{fig7a}}
\hspace{\uhoriz em}
\subfloat[]{\includegraphics[width= \uwidth in]{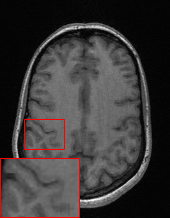}
\label{fig7b}}
\hspace{\uhoriz em}
\subfloat[]{\includegraphics[width= \uwidth in]{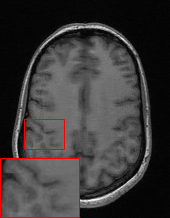}
\label{fig7c}}
\hspace{\uhoriz em}
\subfloat[]{\includegraphics[width= \uwidth in]{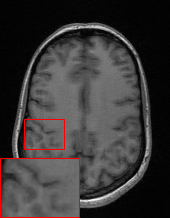}
\label{fig7d}}
\hspace{\uhoriz em}
\subfloat[]{\includegraphics[width= \uwidth in]{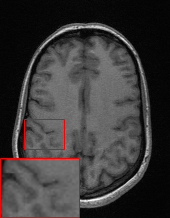}
\label{fig7e}}
\hspace{\uhoriz em}
\subfloat[]{\includegraphics[width= \uwidth in]{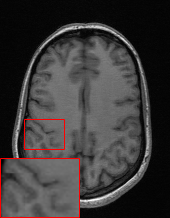}
\label{fig7f}}
\\
\vspace{-0.5 em}
\subfloat[]{\includegraphics[width= \uwidth in]{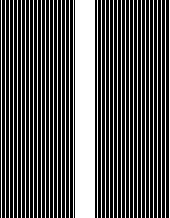}
\label{fig7g}}
\hspace{\uhoriz em}
\subfloat[]{\includegraphics[width= \uwidth in]{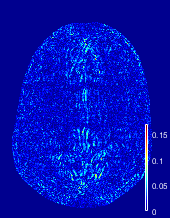}
\label{fig7h}}
\hspace{\uhoriz em}
\subfloat[]{\includegraphics[width= \uwidth in]{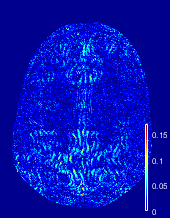}
\label{fig7i}}
\hspace{\uhoriz em}
\subfloat[]{\includegraphics[width= \uwidth in]{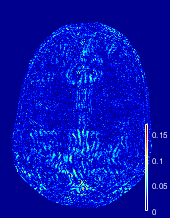}
\label{fig7j}}
\hspace{\uhoriz em}
\subfloat[]{\includegraphics[width= \uwidth in]{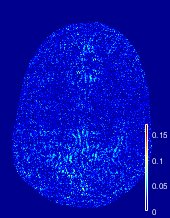}
\label{fig7k}}
\hspace{\uhoriz em}
\subfloat[]{\includegraphics[width= \uwidth in]{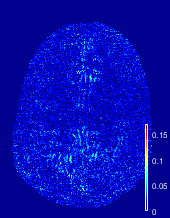}
\label{fig7l}}
}
\end{minipage}
}
\vspace{-0.2em} 
\caption{Reconstructed results of dataset 1 based on the 1DUU pattern $(AF=3)$. (a) Reference image; (b) reconstructed image via JTV-SPIRiT \cite{Weller2014}; (c) reconstructed image via LPJTV-ESPIRiT \cite{Duan2019}; (d) reconstructed image via STDLR-SPIRiT \cite{Zhang2020}; (e) reconstructed image via the NLR-SPIRiT-baseline; (f) reconstructed image by NLR-SPIRiT; (g) undersampling pattern; (h), (i), (j), (k), and (l) show the error maps of (b), (c), (d), (e), and (f), respectively.}
\vspace{-0.5em} 
\label{fig7}
}
\end{figure*}

\begin{figure*}[!t]
\footnotesize{
\centering{ %

\newcommand{\uwidth}{0.85}   
\newcommand{\uhoriz}{-0.5}  

\begin{minipage}[b]{ 0.976 \textwidth} %
\centering{
\subfloat[]{\includegraphics[width= \uwidth in]{data2_enlarge_I1_R5}
\label{fig8a}}
\hspace{\uhoriz em}
\subfloat[]{\includegraphics[width= \uwidth in]{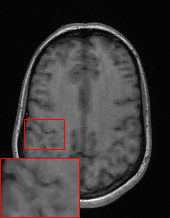}
\label{fig8b}}
\hspace{\uhoriz em}
\subfloat[]{\includegraphics[width= \uwidth in]{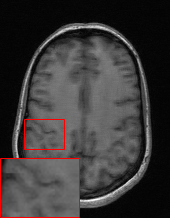}
\label{fig8c}}
\hspace{\uhoriz em}
\subfloat[]{\includegraphics[width= \uwidth in]{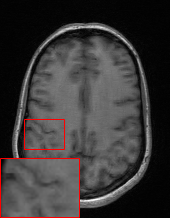}
\label{fig8d}}
\hspace{\uhoriz em}
\subfloat[]{\includegraphics[width= \uwidth in]{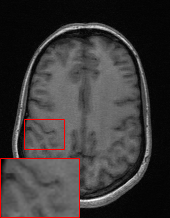}
\label{fig8e}}
\hspace{\uhoriz em}
\subfloat[]{\includegraphics[width= \uwidth in]{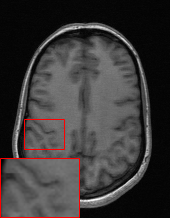}
\label{fig8f}}
\\
\vspace{-0.5 em}
\subfloat[]{\includegraphics[width= \uwidth in]{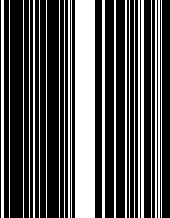}
\label{fig8g}}
\hspace{\uhoriz em}
\subfloat[]{\includegraphics[width= \uwidth in]{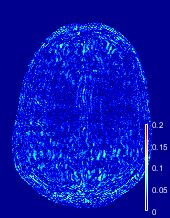}
\label{fig8h}}
\hspace{\uhoriz em}
\subfloat[]{\includegraphics[width= \uwidth in]{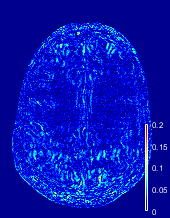}
\label{fig8i}}
\hspace{\uhoriz em}
\subfloat[]{\includegraphics[width= \uwidth in]{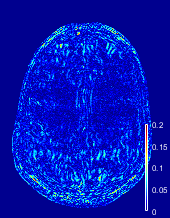}
\label{fig8j}}
\hspace{\uhoriz em}
\subfloat[]{\includegraphics[width= \uwidth in]{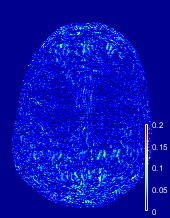}
\label{fig8k}}
\hspace{\uhoriz em}
\subfloat[]{\includegraphics[width= \uwidth in]{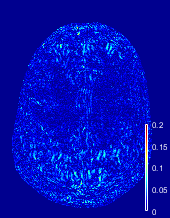}
\label{fig8l}}
}
\end{minipage}
}
\vspace{-0.2em} 
\caption{Reconstructed results of dataset 1 based on the 1DGU pattern $(AF=3)$. (a) Reference image; (b) reconstructed image via JTV-SPIRiT \cite{Weller2014}; (c) reconstructed image via LPJTV-ESPIRiT \cite{Duan2019}; (d) reconstructed image via STDLR-SPIRiT \cite{Zhang2020}; (e) reconstructed image via the NLR-SPIRiT-baseline; (f) reconstructed image by NLR-SPIRiT; (g) undersampling pattern; (h), (i), (j), (k), and (l) show the error maps of (b), (c), (d), (e), and (f), respectively.}
\vspace{-0.5em} 
\label{fig8}
}
\end{figure*}

We also visually compare the images via the proposed NLR-SPIRiT and all considered algorithms for dataset 1 based on the 1DGU pattern and 1DUU pattern ($AF=3$), as shown in Fig. \ref{fig7} and Fig. \ref{fig8}. The NLR-SPIRiT algorithm achieves the least artifacts inside the image and the least errors at the edge, so the NLR-SPIRiT algorithm yields the highest visual quality among all considered algorithms.

\begin{table}[!h]
\centering
\footnotesize{
\renewcommand{\arraystretch}{1}  
\addtolength{\tabcolsep}{-4.2pt}    
\caption{Comparison of the three metrics of the reconstructed images via the competing algorithms based on 1DUU and 1DGU patterns $(AF=3-6)$ for dataset 1.}
    \begin{tabular}{cccccccccc}
    \hline
    \hline
    \multirow{2}[4]{*}{Algorithm} & \multirow{2}[4]{*}{mertic} & \multicolumn{4}{c}{1DUU}      & \multicolumn{4}{c}{1DGU} \\
    \cline{3-10}          &       & AF=3   & AF=4   & AF=5   & AF=6   & AF=3   & AF=4   & AF=5   & AF=6 \\
    \hline
    \multirow{3}[2]{*}{JTV-SPIRiT \cite{Weller2014}} & SNR   & 21.71 & 19.48 & 17.06 & 16.56 & 19.19 & 16.89 & 15.52 & 13.55 \\
          & HFEN  & 0.155 & 0.2089 & 0.2956 & 0.3067 & 0.2132 & 0.2877 & 0.3443 & 0.4412 \\
          & SSIM  & 0.9389 & 0.9129 & 0.8765 & 0.8723 & 0.9173 & 0.8828 & 0.8635 & 0.833 \\
    \hline
    \multirow{3}[2]{*}{STDLR-SPIRiT \cite{Zhang2020}} & SNR   & 21.11 & 19.16 & 17.7  & 17.09 & 19.21 & 17.04 & 15.84 & 13.7 \\
          & HFEN  & 0.1815 & 0.2304 & 0.2755 & 0.2933 & 0.219 & 0.2896 & 0.3368 & 0.4434 \\
          & SSIM  & 0.9468 & 0.9247 & 0.9059 & 0.8984 & 0.9338 & 0.9047 & 0.8915 & 0.8655 \\
    \hline
    \multirow{3}[2]{*}{pFISTA-SPIRiT \cite{Zhang2021}} & SNR   & 21.19 & 19.18 & 17.14 & 15.98 & 18.89 & 16.4  & 14.85 & 12.65 \\
          & HFEN  & 0.174 & 0.223 & 0.2906 & 0.3339 & 0.2199 & 0.3055 & 0.3794 & 0.5026 \\
          & SSIM  & 0.9496 & 0.9286 & 0.9035 & 0.8926 & 0.9357 & 0.904 & 0.884 & 0.8522 \\
    \hline
    \multirow{3}[2]{*}{NLR-SPIRiT-baseline} & SNR   & 22.37 & 20.53 & 18.87 & 17.81 & 20.31 & 18.12 & 16.35 & 14.11 \\
          & HFEN  & 0.1392 & 0.1874 & 0.2347 & 0.2598 & 0.1836 & 0.2483 & 0.3119 & 0.418 \\
          & SSIM  & 0.9555 & 0.9401 & 0.9197 & 0.9103 & 0.9426 & 0.918 & 0.9001 & 0.8746 \\
    \hline
    \multirow{3}[2]{*}{NLR-SPIRiT} & SNR   & \textbf{23.16} & \textbf{20.79} & \textbf{19.2} & \textbf{18.58} & \textbf{21.02} & \textbf{18.7} & \textbf{17.26} & \textbf{15.16} \\
          & HFEN  & \textbf{0.1289} & \textbf{0.1873} & \textbf{0.2306} & \textbf{0.2453} & \textbf{0.1738} & \textbf{0.2367} & \textbf{0.2789} & \textbf{0.364} \\
          & SSIM  & \textbf{0.9633} & \textbf{0.9425} & \textbf{0.9239} & \textbf{0.9156} & \textbf{0.9489} & \textbf{0.9212} & \textbf{0.9072} & \textbf{0.8846} \\
    \hline
	\hline
    \end{tabular}%
  \label{tab5}%
  \vspace{-1em}
  }
\end{table}%

Table \ref{tab5} summarizes the SNR, HFEN, and SSIM values of the reconstructed images shown in Fig. \ref{fig7} and Fig. \ref{fig8}. As indicated in Table \ref{tab5} ,NLR-SPIRiT achieves SNR improvements of 1.74 dB, 1.63 dB, 2.21 dB, and 0.68 dB on average over JTV-SPIRiT, STDLR-SPIRiT, pFISTA-SPIRiT, and NLR-SPIRiT-baseline, respectively. In a word, the proposed NLR-SPIRiT can more effectively reconstruct the MR images from undersampled k-space data based on 1D undersampling patterns.

\begin{figure}[!t]
\footnotesize{
\centering{ %

\newcommand{\uwidth}{1.7} 
\newcommand{\uhoriz}{-0.6}  

\begin{minipage}[b]{ 0.976 \textwidth} %
\centering{
\subfloat[]{\includegraphics[width= \uwidth in]{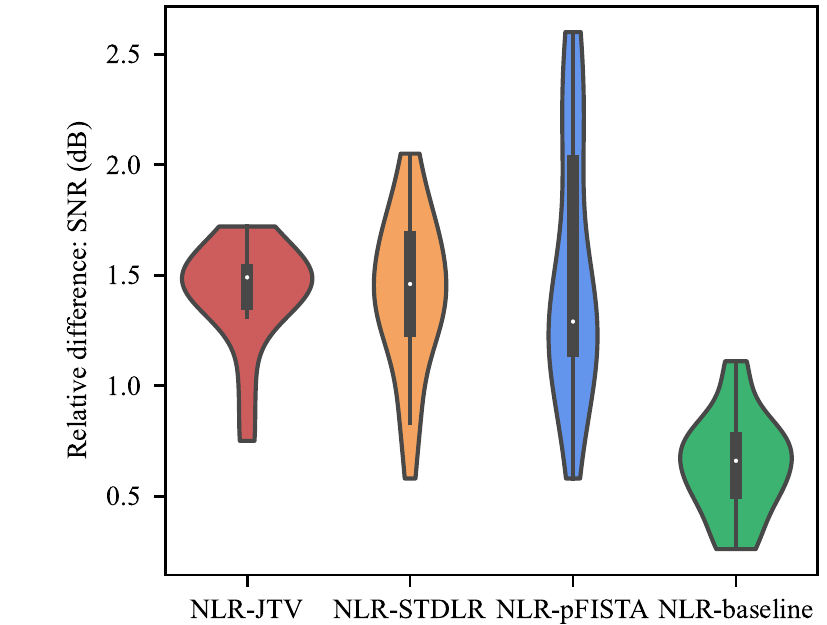}
\label{fig6a}}
\hspace{\uhoriz em}
\subfloat[]{\includegraphics[width= \uwidth in]{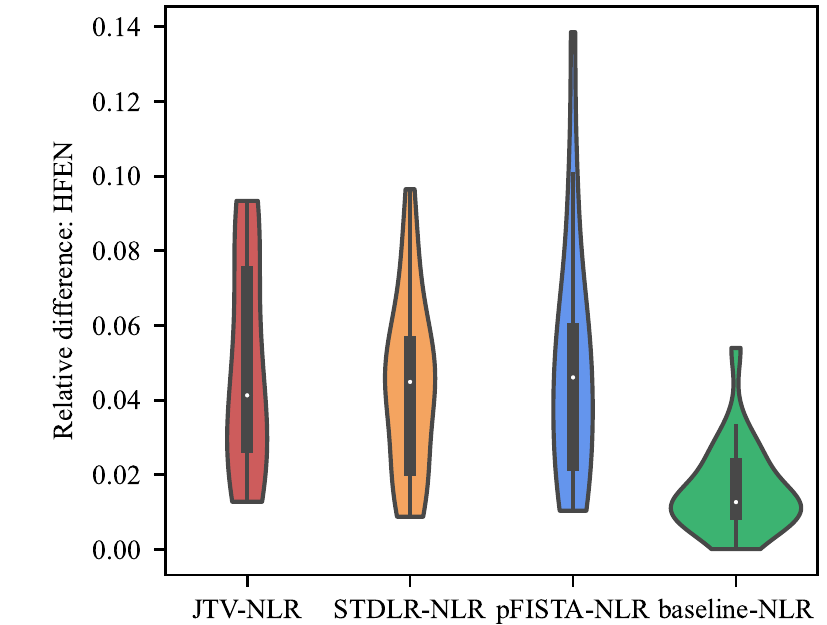}
\label{fig6b}}
\hspace{\uhoriz em}
\subfloat[]{\includegraphics[width= \uwidth in]{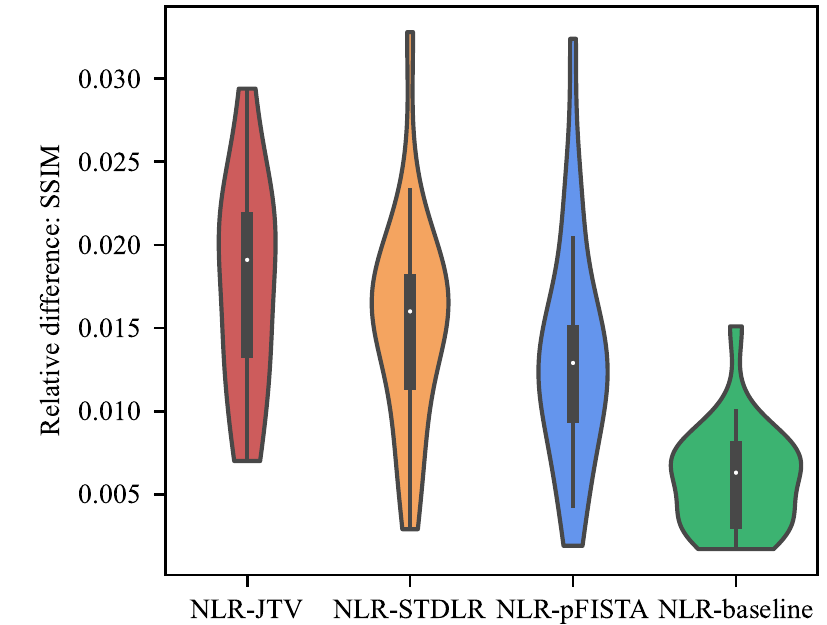}
\label{fig6c}}
}
\end{minipage}
}
\vspace{-0.8em} 
\caption{Violin plots show the relative differences of (a) SNR, (b) HFEN, and (c) SSIM between NLR-SPIRiT and JTV-SPIRiT, STDLR-SPIRiT, pFISTA-SPIRiT and NLR-SPIRiT-baseline, which are simplified as NLR, JTV, STDLR, pFISTA, and baseline in plots, respectively.}
\vspace{-0.2em} 
\label{fig6}
}
\end{figure}

We also provide violin plots for the relative metric differences between NLR-SPIRiT and the comparing algorithms in Tables \ref{tab2}-\ref{tab5}. As shown in Fig. \ref{fig6}, NLR-SPIRiT significantly outperforms all comparing algorithms in terms of SNR, HFEN, and SSIM.

\subsubsection{Comparison results of NLR-SPIRiT-baseline and NLR-SPIRiT}

The NLR-SPIRiT algorithm and the NLR-SPIRiT-baseline algorithm use different surrogate functions to solve the LR minimization problem, corresponding to the WNN and the NN, respectively.

The previous experimental results has compared JTV-SPIRiT, STDLR-SPIRiT, pFISTA-SPIRiT, NLR-SPIRiT-baseline and NLR-SPIRiT algorithms. As shown in Figs. \ref{fig3}-\ref{fig6} and Tables \ref{tab2}-\ref{tab5}, NLR-SPIRiT-baseline almost outperforms all other considered algorithms for dataset 1-3 based on 2DPU patterns and two different 1D undersampling patterns in terms of visual comparison and three metrics. NLR-SPIRiT performs better than NLR-SPIRiT-baseline, and outperforms other considered algorithms.

In other words, the introduction of the NLR regularization term contributes to the reconstruction performance improvement of these SPIRiT-based algorithms. Furthermore, using the WNN can achieve better visual performance and numerical metrics than using the NN.

\subsubsection{Further discussion about performance of the proposed algorithm}
\begin{figure*}[!t]
\footnotesize{
\centering{ %

\newcommand{\uwidth}{0.85}   
\newcommand{\uhoriz}{-0.5}  

\begin{minipage}[b]{ 0.976 \textwidth} %
\centering{
\subfloat[]{\includegraphics[width= \uwidth in]{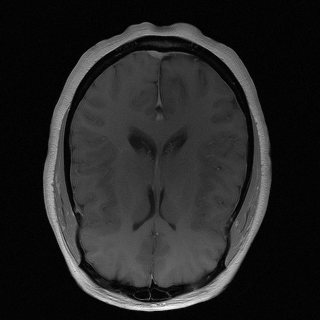}
\label{fig10a}}
\hspace{\uhoriz em}
\subfloat[]{\includegraphics[width= \uwidth in]{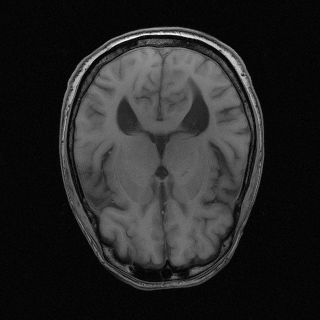}
\label{fig10b}}
\hspace{\uhoriz em}
\subfloat[]{\includegraphics[width= \uwidth in]{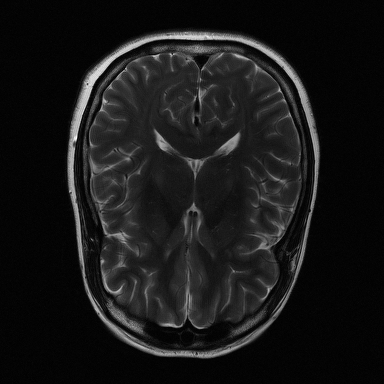}
\label{fig10c}}
\hspace{\uhoriz em}
\subfloat[]{\includegraphics[width= \uwidth in]{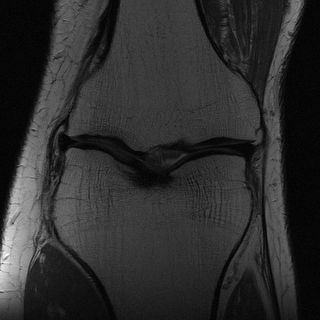}
\label{fig10d}}
\hspace{\uhoriz em}
\subfloat[]{\includegraphics[width= \uwidth in]{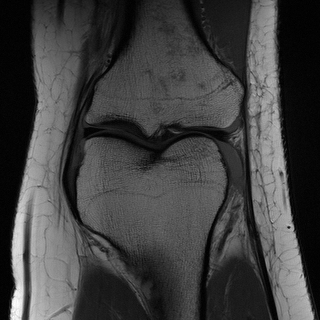}
\label{fig10e}}
\hspace{\uhoriz em}
\subfloat[]{\includegraphics[width= \uwidth in]{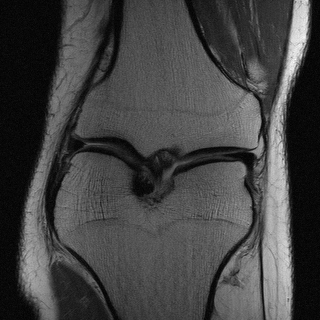}
\label{fig10f}}
}
\end{minipage}
}
\vspace{-0.2em} 
\caption{Six MR images used for further discussing performance of the proposed algorithm.}
\vspace{-0.5em} 
\label{fig10}
}
\end{figure*}

\begin{table}[!t]
\centering
\footnotesize{
\renewcommand{\arraystretch}{1}  
\addtolength{\tabcolsep}{-1pt}    
\caption{The average SNR, HFEN and SSIM differences between NLR-SPIRiT and the considered SPIRiT-based algorithms for reconstructing six data sets shown in Fig. \ref{fig10} based on the 2DPU patterns with $AF=3-7$.}
    \begin{tabular}{ccccccc}
    \hline
    \hline
        Competing algorithms & mertic & AF=3  & AF=4  & AF=5  & AF=6  & AF=7 \\
    \hline
    \multirow{3}[2]{*}{JTV-SPIRiT \cite{Weller2014}} & SNR   & 0.52  & 0.62  & 0.83  & 0.97  & 1.19  \\
          & HFEN  & 0.0148  & 0.0205  & 0.0283  & 0.0329  & 0.0409  \\
          & SSIM  & 0.0233  & 0.0303  & 0.0361  & 0.0419  & 0.0481  \\
    \hline
    \multirow{3}[2]{*}{STDLR-SPIRiT \cite{Zhang2020}} & SNR   & 0.74  & 0.97  & 1.21  & 1.37  & 1.55  \\
          & HFEN  & 0.0158  & 0.0258  & 0.0372  & 0.0442  & 0.0518  \\
          & SSIM  & 0.0051  & 0.0076  & 0.0103  & 0.0126  & 0.0151  \\
    \hline
    \multirow{3}[2]{*}{pFISTA-SPIRiT \cite{Zhang2021}} & SNR   & 0.81  & 0.81  & 0.97  & 1.09  & 1.32  \\
          & HFEN  & 0.0078  & 0.0112  & 0.0189  & 0.0246  & 0.0366  \\
          & SSIM  & 0.0046  & 0.0044  & 0.0051  & 0.0073  & 0.0100  \\
    \hline
	\hline
    \end{tabular}%
  \label{tab7}%
  \vspace{-1em}
  }
\end{table}%

Another six datasets are used to further verify the image reconstruction performance of the proposed algorithm, including three brain datasets and three knee datasets, as shown in Fig. \ref{fig10}. These datasets are multi-coil raw k-space data from NYU fastMRI \cite{Zbontar2018,Knoll2020}. The three brain datasets come from the first frame of the dataset ``brain\_AXT1POST\_200\_6001959", ``brain\_AXT1\_201\_6002688" and ``brain\_AXT2\_200\_2000003" acquired by using a 20-channel coil (matrix size = $320\times320$). In our experiments, we use the coil compression technique \cite{Bahri2013} to compress these data from 15-channel to 8-channel. The three knee datasets are the 20th slice of the dataset ``knee\_file1000005", ``knee\_file1000010", and ``knee\_file1000012" acquired on a 15-channel coil (matrix size = $320\times640$). These knee datasets are cropped into the size of $320\times320$ and compressed into eight channels in the experiments.

The NLR-SPIRiT algorithm and previous SPIRiT-based algorithms are used to reconstruct six datasets based on 2DPU patterns and $AF=3-7$. For different AF values, the average differences of the three metrics between the NLR-SPIRiT algorithm and other compared algorithms are calculated and shown in Table \ref{tab7}.

Table \ref{tab7} shows that the STDLR-SPIRiT algorithm reconstructs the images with the lowest SNR and SSIM. pFISTA-SPIRiT is better than STDLR-SPIRiT. The JTV-SPIRiT algorithm outperforms pFISTA-SPIRiT. And the NLR-SPIRiT algorithm has a noticeable improvement over other compared algorithms in all three metrics.

\begin{table}[!t]
\centering
\footnotesize{
\renewcommand{\arraystretch}{1}  
\addtolength{\tabcolsep}{-0pt}    
\caption{Comparison of the runtime and memory demands of the considered PMRI reconstruction algorithm for the dataset 1 based on the 2DPU pattern $(AF=5)$.}
    \begin{tabular}{cccccccc}
    \hline
    \hline
    Algorithms & runime (s) & memory demand (GB) \\
    \hline
    {JTV-SPIRiT \cite{Weller2014}}  & 28    & 0.46 \\
    {STDLR-SPIRiT \cite{Zhang2020}}  & 1273  & 45.35 \\
    {pFISTA-SPIRiT \cite{Zhang2021}}  & 181   & 0.50 \\
    {NLR-SPIRiT} & 796   & 0.49 \\
    \hline
    \hline
    \end{tabular}%
  \label{tab6}%
  \vspace{-1em}
  }
\end{table}%

\subsubsection{Comparison of runtime and memory demand}
Next, we compare the runtime and memory demands of JTV-SPIRiT, STDLR-SPIRiT, pFISTA-SPIRiT, and NLR-SPIRiT. Table \ref{tab6} summarizes the runtime and memory demands of the considered algorithms in PMRI reconstruction for dataset 1 based on the 2DPU pattern ($AF=5$). As indicated in Table \ref{tab6}, JTV-SPIRiT has advantages in runtime and memory demands. STDLR-SPIRiT has the highest computational complexity, requires the longest runtime, and the maximum memory. Compared with STDLR-SPIRiT, NLR-SPIRiT requires 0.49 GB of RAM memory for reconstruction, only approximately one-92th of the memory demand of STDLR-SPIRiT. Overall, the required runtime of NLR-SPIRiT is more than JTV-SPIRiT and pFISTA-SPIRiT but less than STDLR-SPIRiT. And the required memory of STDLR-SPIRiT is more than other SPIRiT-based method.

In fact, the proposed NLR-SPIRiT method has a lot of improvement room in runtime. The runtime of NLR-SPIRiT is $K \times ({{{t_{BM}}} \mathord{\left/{\vphantom {{{t_{BM}}} T}} \right. \kern-\nulldelimiterspace} T} + {t_{NLR}} + {t_1}) + {t_0}$, where $K$ is the outer iteration number, ${t_{BM}}$ is the runtime of the block-matching operation performed every $T$ iterations, ${t_{NLR}}$ is the runtime of one LR approximation, ${t_1}$ is the runtime to update $Z$, $X$  and ${u_Z}$ once, and ${t_0}$ is the runtime of algorithm initialization and preprocessing. The very high computational complexity of the LR approximation and block-matching steps result in large ${t_{NLR}}$ and ${t_{BM}}$ values. Therefore, we will optimize the LR approximation step and block-matching steps with a graphics processing unit (GPU) to efficiently accelerate the proposed NLR-SPIRiT algorithm in the future.

\begin{table}[!t]
\centering
\footnotesize{
\renewcommand{\arraystretch}{1}  
\addtolength{\tabcolsep}{-3pt}    
\caption{ SNR values for reconstructing undersampled dataset 1 based on the 2DPU pattern ($AF=5$) when using different autocalibration signal (ACS) sizes and calibration kernel sizes (KS).}
    \begin{tabular}{ccccccccccccc}
    \hline
    \hline
    \multicolumn{2}{c}{\multirow{2}[4]{*}{SNR}} & \multicolumn{11}{c}{ACS size} \\
    \cline{3-13}   & & 6$\times$6     & 7$\times$7     & 8$\times$8    & 9$\times$9     & 10$\times$10   & 11$\times$11    & 12$\times$12    & 13$\times$13    & 14$\times$14   & 20$\times$20    & 24$\times$24 \\
    \hline
    \multirow{3}[2]{*}{KS} & 3$\times$3    & 14.33 & 20.44 & 21.07 & 21.57 & 21.7 & 21.81 & 21.85 & 21.97 & 22.03  & 22.43 & 22.68 \\
          & 5$\times$5     & -     & -     & -     & 13.63  & 21.01 & 22.1   & 22.21 & 22.26 & 22.34  & 22.75 & 22.98 \\
          & 7$\times$7     & -     & -     & -     & -     & -    & -     & 12.76 & 19.58 & 22.14  & 22.71 & 22.99 \\
    \hline
	\hline
    \end{tabular}%
  \label{tab1}%
  \vspace{-1em}
  }
\end{table}%

\subsection{Comparison to different ACS size and calibration kernel size}
In the calibration consistency step of the SPIRiT-based algorithm, the ACS size and the calibration kernel size (KS) are vital parameters. The experiment tests the performance of the proposed NLR-SPIRiT algorithm when using different ACS sizes and ${\text{KS}} = 3 \times 3,5 \times 5,7 \times 7$ in 2DPU patterns. Table \ref{tab1} shows the SNR values for reconstructing undersampled dataset 1 based on the 2DPU pattern ($AF=5$).

As shown in Table \ref{tab1},  the experimental results reveal: a) the small ACS region will degrade the reconstruction image quality to some extent, so we chose the ACS size as $24 \times 24$ for a better reconstruction quality; b) when the ACS size is large enough (such as $ACS>10\times10$), the SNR with ${\text{KS}} =5\times5$ is higher than that with ${\text{KS}} =3\times3$ and ${\text{KS}} =7\times7$. So we chose KS as 5×5 for a better reconstruction quality; c) for some smaller ACS sizes ($ACS=7\times7$ - $10\times10$), we can reduce KS (such as ${\text{KS}} =3\times3$) to guarantee the image reconstruction quality. In addition, the SNR dramatically drops to a low value (14.33 dB) when ${\text{KS}} =3\times3$, with the ACS size reducing to $6\times6$. At this time, NLR-SPIRiT cannot reconstruct images well.



\begin{figure}[!t]
\footnotesize{
\centering{ %
\newcommand{\uwidth}{2.5} 
\newcommand{\uhoriz}{-0}  
\begin{minipage}[b]{ 0.976 \textwidth} %
\centering{
\includegraphics[width= \uwidth in]{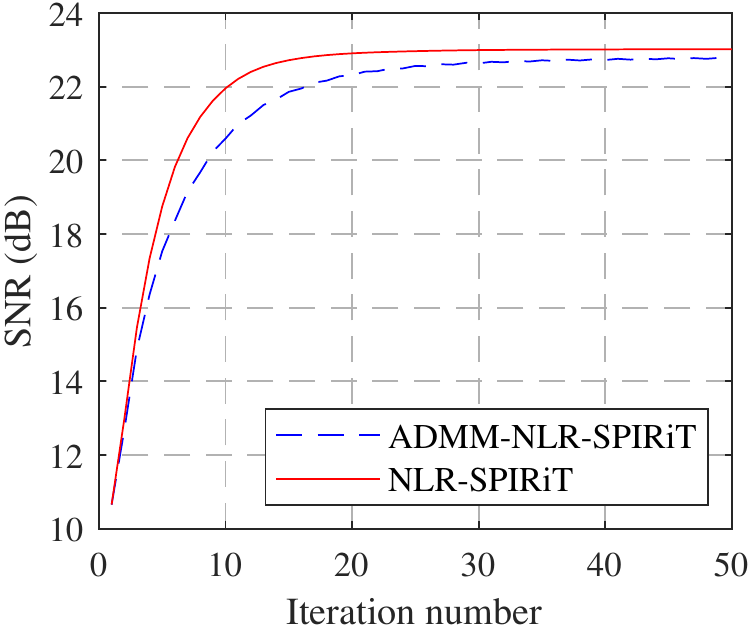}
}
\end{minipage}
}
\vspace{-1em} 
\caption{SNR curves of ADMM-NLR-SPIRiT and NLR-SPIRiT for dataset 1 based on the 2DPU pattern $(AF=5)$.}
\vspace{-0.5em} 
\label{fig9}}
\end{figure}

\subsection{Proposed NE formulation versus ADMM formulation}

According to the derivations of the NLR-SPIRiT model and ADMM-NLR-SPIRiT algorithm (see the appendix), we observe that these two algorithms attain almost the same calculation time for one-iteration.

We compared the reconstruction results obtained with NLR-SPIRiT and ADMM-NLR-SPIRiT for dataset 1 based on the 2DPU pattern $(AF=5)$. As shown in Fig. \ref{fig9}, NLR-SPIRiT attains faster convergence and obtains a slightly higher SNR value than ADMM-NLR-SPIRiT. And we determine that they require almost the same reconstruction time every iteration. In addition, NLR-SPIRiT contains two fewer parameters than does ADMM-NLR-SPIRiT, and therefore it is easier to tune the parameters of NLR-SPIRiT. In summary, it is concluded that the proposed NLR-SPIRiT algorithm exhibits obvious advantages in practical applications.

\subsection{Further discussion about parameter settings}
\begin{figure}[!t]
\footnotesize{
\centering{ %

\newcommand{\uwidth}{2.5} 
\newcommand{\uhoriz}{-0.6}  

\begin{minipage}[b]{ 0.976 \textwidth} %
\centering{
\subfloat[]{\includegraphics[width= \uwidth in]{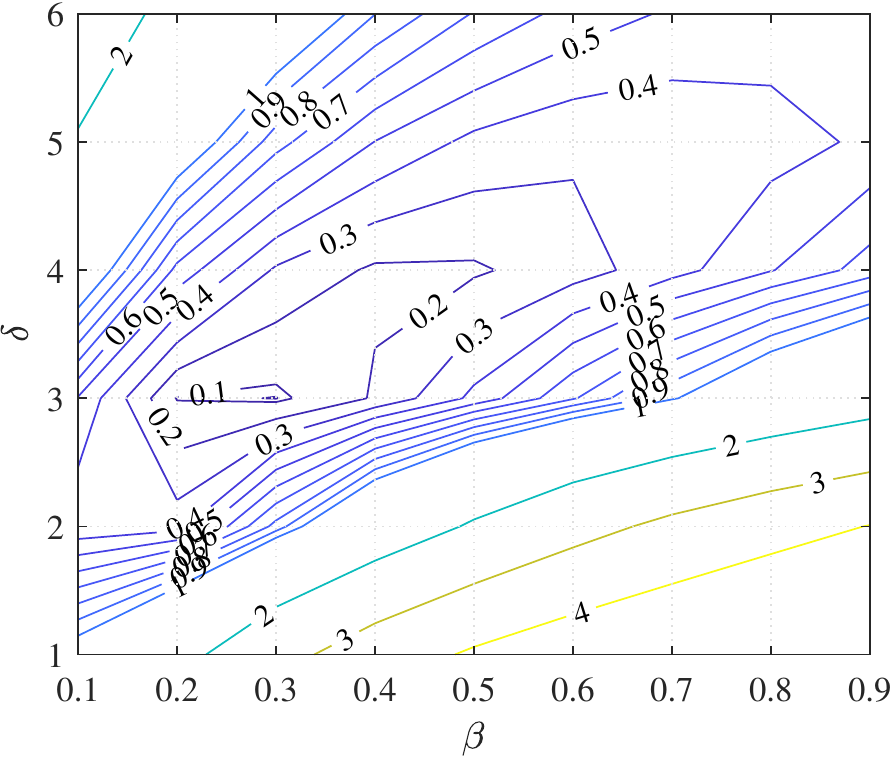}
\label{fig11a}}
\hspace{\uhoriz em}
\subfloat[]{\includegraphics[width= \uwidth in]{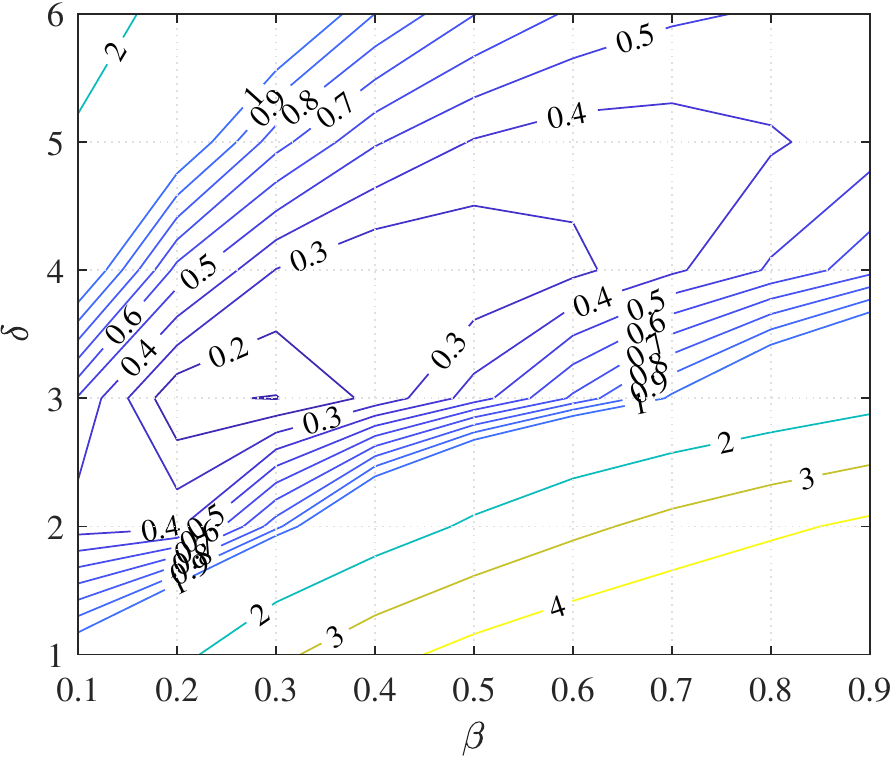}
\label{fig11b}}
}
\end{minipage}
}
\vspace{-0.8em} 
\caption{We ran NLR-SPIRiT on the 15 training and 32 validation undersampling data sets on the ranges of $\delta  \in [1,6]$ and $\beta  \in [0.1,1]$, and obtained the average SNR relative differences between the maximum SNR and the SNR for each parameter setting. Contour plots show the average SNR relative differences of (a) 15 training data sets and (b) 32 validation data sets. The average SNR relative differences are also minimized around $\delta=3$ and $\beta=0.3$.}
\vspace{-0.2em} 
\label{fig11}
}
\end{figure}

We found that there exist similar parameters for reconstructing the MR images of same kind, so we also offer a parameter selection method for practical application. First, we train the parameters from 15 retrospective undersampling data sets with the 2DPU pattern ($AF = 5$). We ran NLR-SPIRiT on the training undersampling data sets on the ranges of   $\delta  \in [1,6]$ and $\beta  \in [0.1,1]$, and obtained SNRs for different parameter settings. The average SNR relative difference between the maximum SNR and the SNR for each parameter setting is calculated. As shown in Fig. \ref{fig11}(a), the average SNR relative differences are minimized at $\delta=3$ and $\beta=0.3$. Second, we ran NLR-SPIRiT on the 32 retrospective validation undersampling data sets on the ranges of $\delta  \in [1,6]$ and $\beta  \in [0.1,1]$, and obtained the average SNR relative differences for each parameter setting. As shown in Fig. \ref{fig11}(b), the average SNR relative differences are also minimized around $\delta=3$ and $\beta=0.3$. Therefore, for this kind of data sets, the optimal parameter setting was determined to be $\delta=3$ and $\beta=0.3$.

\begin{figure*}[!t]
\footnotesize{
\centering{ %

\newcommand{\uwidth}{0.85}   
\newcommand{\uhoriz}{-0.5}  

\begin{minipage}[b]{ 0.976 \textwidth} %
\centering{
\subfloat[]{\includegraphics[width= \uwidth in]{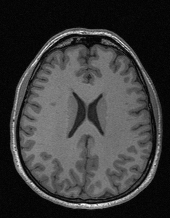}
\label{fig12a}}
\hspace{\uhoriz em}
\subfloat[]{\includegraphics[width= \uwidth in]{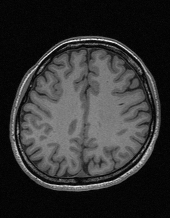}
\label{fig12b}}
\hspace{\uhoriz em}
\subfloat[]{\includegraphics[width= \uwidth in]{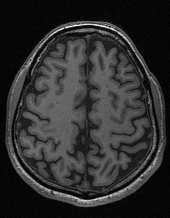}
\label{fig12c}}
\hspace{\uhoriz em}
\subfloat[]{\includegraphics[width= \uwidth in]{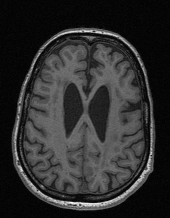}
\label{fig12d}}
\hspace{\uhoriz em}
\subfloat[]{\includegraphics[width= \uwidth in]{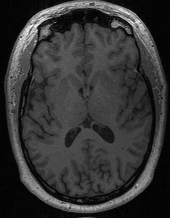}
\label{fig12e}}
\hspace{\uhoriz em}
\subfloat[]{\includegraphics[width= \uwidth in]{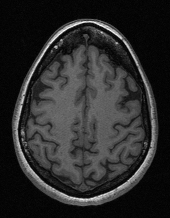}
\label{fig12f}}
}
\end{minipage}
}
\vspace{-0.2em} 
\caption{six MR images of same kind with dataset 1.}
\vspace{-0.5em} 
\label{fig12}
}
\end{figure*}

\begin{table}[!t]
\centering
\footnotesize{
\renewcommand{\arraystretch}{1}  
\addtolength{\tabcolsep}{-1pt}    
\caption{The average SNR, HFEN and SSIM differences between NLR-SPIRiT and the considered SPIRiT-based algorithms for reconstructing six data sets shown in Fig. \ref{fig12} based on the 2DPU patterns with $AF=3-7$.}
    \begin{tabular}{ccccccc}
    \hline
    \hline
    Competing algorithms& mertic & \multicolumn{1}{l}{AF=3} & \multicolumn{1}{l}{AF=4} & \multicolumn{1}{l}{AF=5} & \multicolumn{1}{l}{AF=6} & \multicolumn{1}{l}{AF=7} \\
    \hline
    \multirow{3}[2]{*}{JTV-SPIRiT \cite{Weller2014}} & SNR   & 0.76  & 1.02  & 1.15  & 1.23  & 1.32  \\
          & HFEN  & 0.0110  & 0.0170  & 0.0207  & 0.0246  & 0.0288  \\
          & SSIM  & 0.0132  & 0.0188  & 0.0226  & 0.0269  & 0.0296  \\
    \hline
    \multirow{3}[2]{*}{STDLR-SPIRiT \cite{Zhang2020}} & SNR   & 0.79  & 1.33  & 1.60  & 1.71  & 1.68  \\
          & HFEN  & 0.0161  & 0.0297  & 0.0387  & 0.0449  & 0.0452  \\
          & SSIM  & 0.0039  & 0.0102  & 0.0151  & 0.0200  & 0.0212  \\
    \hline
    \multirow{3}[2]{*}{pFISTA-SPIRiT \cite{Zhang2021}} & SNR   & 0.72  & 1.06  & 1.18  & 1.18  & 1.15  \\
          & HFEN  & 0.0092  & 0.0157  & 0.0195  & 0.0197  & 0.0212  \\
          & SSIM  & 0.0028  & 0.0071  & 0.0095  & 0.0114  & 0.0108  \\
    \hline
	\hline
    \end{tabular}%
  \label{tab8}%
  \vspace{-1em}
  }
\end{table}%

We also compare NLR-SPIRiT with the parameters $\delta=3$, $\beta=0.3$ and the competing algorithms with optimal parameters. Table \ref{tab8} tabulates the average SNR, HFEN and SSIM differences between NLR-SPIRiT and the competing algorithms for reconstructing six data sets shown in Fig. \ref{fig12} based on the 2DPU patterns with $AF=3-7$. As shown in Table \ref{tab8}, NLR-SPIRiT with the parameters $\delta=3$, $\beta=0.3$  can achieve better reconstruction performance than the competing algorithms with optimal parameters in terms of SNR, HFEN and SSIM.

\section{Conclusions}
\label{conclusion}

In this paper, we propose the NLR-SPIRiT model, which incorporates the NLR regularization into the SPIRiT model. The NLR-SPIRiT model fully utilizes both the NSS in MR images and the calibration consistency in the k-space domain. We adopt the WNN instead of the NN as a surrogate of the rank, and employ the NE formulation and the ADMM technique to efficiently solve the NLR-SPIRiT model. Experimental results considering different in vivo datasets and undersampling patterns indicate that the proposed NLR-SPIRiT algorithm almost achieves a better performance in terms of three objective metrics and visual perception over state-of-the-art methods. In addition, we propose a parameter setting method for practical application, which selects a set of near-optimal parameters for the same kind of MR images. We will optimize the most time-consuming BM and LR approximation steps with a GPU to efficiently accelerate the NLR-SPIRiT algorithm in the future. The proposed NLR-SPIRiT algorithm is very promising in regard to PMRI applications.

\section*{Acknowledgment}
The authors would like to acknowledge Michael Lustig, Daniel S. Weller, Xinlin Zhang, Kevin Epperson, Fa-Hsuan Lin, Robreto Souza, Saiprasad Ravishankar, Bihan Wen, and Weisheng Dong, for making their code or in vivo data sets publicly available. The authors would like to thank the anonymous reviewers for their comments and suggestions.

\appendix
\section{Alternative algorithm}
\label{appendix}

The PMRI reconstruction problem based on the NLR regularization and the SPIRiT model can be rewritten as follows:
\begin{equation}
\begin{aligned}
X = \arg \;\mathop {\min }\limits_X  \frac{1}{2}\left\| {AX - Y} \right\|_F^2 + \frac{{{\mu _1}}}{2}\left\| {(G - I)X} \right\|_F^2 + \tau \sum\limits_{c = 1}^C {\sum\limits_{i = 1}^{{N_p}} {{\mathop{\rm rank}\nolimits} \left( {{V_{ci}}(X)} \right)} } \label{eq10b}
\end{aligned}
\end{equation}

Problem (\ref{eq10b}) can be solved with the variable splitting (VS) and ADMM techniques. First of all, by introducing auxiliary variables  $Z = X$,  ${D_{ci}} = {V_{ci}}\left( X \right)$,  $B = X$ and  $D = [{D_{11}},...,{D_{ci}},...,{D_{C{N_p}}}]$, in addition to corresponding Lagrange multipliers  ${u_Z}$,  ${u_{{D_{ci}}}}$,  ${u_B}$, and  ${u_D} = [{u_{{D_{11}}}},...,{u_{{D_{ci}}}},...,{u_{{D_{C{N_p}}}}}]$, respectively,  problem (\ref{eq10b}) is decomposed into the following subproblems via the ADMM method:
\begin{equation}
{Z^{k + 1}} = \arg \mathop {\min }\limits_Z {\mu _1}\left\| {\left( {G - I} \right)Z} \right\|_F^2 + {\beta _1}\left\| {Z - \left( {{X^k} + u_Z^k} \right)} \right\|_F^2 \label{eq24}
\end{equation}
\begin{equation}
\left\{ {D_{ci}^{k + 1}} \right\} \!=\! \arg \mathop {\min }\limits_{\left\{ {{D_{ci}}} \right\}} \! \frac{{{\beta _2}}}{2} \! \left\| {{D_{ci}} \!-\! \left( {{V_{ci}}({X^k}) \!+\! u_{{D_{ci}}}^k} \right)} \! \right\|_F^2 + \tau {\mathop{\rm rank}\nolimits} ({D_{ci}})\label{eq25}
\end{equation}
\begin{equation}
\begin{aligned}
{B^{k + 1}} \!=\! \arg \mathop {\min }\limits_B  {\beta _2}\sum\limits_{c = 1}^C {\sum\limits_{i = 1}^{{N_p}} {\left\| {{V_{ci}}(B) \!-\! \left( {D_{ci}^{k + 1} \!-\! u_{{D_{ci}}}^k} \right)} \right\|_F^2} } \!+\! {\beta _3}\left\| {B \!-\! {X^k} \!-\! u_B^k} \right\|_F^2 \label{eq26}
\end{aligned}
\end{equation}
\begin{equation}
\begin{aligned}
{X^{k + 1}} \!=\! \arg \mathop {\min }\limits_X  \left\| {AX \!-\! Y} \right\|_F^2 \!+\! {\beta _1}\left\| {X \!-\! \left( {{Z^{k + 1}} \!-\! u_Z^k} \right)} \right\|_F^2  \!+\! {\beta _3}\left\| {X \!-\! ({B^{k \!+\! 1}} \!-\! u_B^k)} \right\|_F^2 \label{eq27}
\end{aligned}
\end{equation}
\begin{equation}
u_Z^{k + 1} = u_Z^k + {\eta _1}\left( {{X^{k + 1}} - {Z^{k + 1}}} \right) \label{eq28}
\end{equation}
\begin{equation}
u_{{D_{ci}}}^{k{\rm{ + }}1} = u_{{D_{ci}}}^k + {\eta _2}\left( {{V_{ci}}({X^{k + 1}}) - D_{ci}^{k + 1}} \right) \label{eq29}
\end{equation}
\begin{equation}
u_B^{k + 1} = u_B^k + {\eta _3}\left( {{X^{k + 1}} - {B^{k + 1}}} \right) \label{eq30}
\end{equation}

As mentioned in Section III of the manuscript, the subproblems (\ref{eq24}) and (\ref{eq25}) are efficiently solved with respect to  $Z$ and  ${D_{ci}}$, respectively. Subproblem (\ref{eq26}) with respect to  $B$ yields the following closed-form solution:
\begin{equation}
{B^{k + 1}} = \frac{{{\beta _2}\sum\limits_{c = 1}^C {\sum\limits_{i = 1}^{{N_p}} {V_{ci}^{\rm{*}}\left( {D_{ci}^{k + 1} - u_{{D_{ci}}}^k} \right)} }  \!+\! {\beta _3}\left( {{X^k} + u_B^k} \right)}}{{{\beta _2}\sum\limits_{c = 1}^C {\sum\limits_{i = 1}^{{N_p}} {V_{ci}^{\rm{*}}{V_{ci}}}  + {\beta _3}I} }}\label{eq31}
\end{equation}

Subproblem (\ref{eq27}) can be solved by the following formulation:
\begin{equation}
\begin{aligned}
\left( {{A^H}A + {\beta _1}I + {\beta _3}I} \right){X^{k + 1}} ={A^H}Y \!+\! {\beta _1}\left( {{Z^{k + 1}} \!-\! u_Z^k} \right) \!+\! {\beta _3}\left( {{B^{k + 1}} \!-\! u_B^k} \right) \label{eq32}
\end{aligned}
\end{equation}

Substituting $A = \mathcal{P}\mathcal{F}$  and multiplying two-dimensional discrete Fourier transform $\mathcal{F}$ on both side of (\ref{eq32}), we can obtain the following equation:
\begin{equation}
\begin{aligned}
&\mathcal{F}\left( {{\mathcal{F}^H}{\mathcal{P}^H}\mathcal{P}\mathcal{F}{\mathcal{F}^{ - 1}}\mathcal{F} + {\beta _1}{\mathcal{F}^{ - 1}}\mathcal{F} + {\beta _3}{\mathcal{F}^{ - 1}}\mathcal{F}} \right){X^{k + 1}} \\
= &\mathcal{F}\left( {{\mathcal{F}^H}{\mathcal{P}^H}Y \!+\! {\beta _1}\left( {{Z^{k + 1}} \!-\! u_Z^k} \right) \!+\! {\beta _3}\left( {{B^{k + 1}} \!-\! u_B^k} \right)} \right) \label{eq33}
\end{aligned}
\end{equation}

Equation (\ref{eq33}) can be rewritten as:
\begin{equation}
\left( {{\mathcal{P}^H}\mathcal{P} + {\beta _1}I + {\beta _3}I} \right)\mathcal{F}{X^{k + 1}} = {\mathcal{P}^H}Y \!+\! \mathcal{F}\left( {{\beta _1}\left( {{Z^{k + 1}} \!-\! u_Z^k} \right) \!+\! {\beta _3}\left( {{B^{k + 1}} \!-\! u_B^k} \right)} \right) \label{eq34}
\end{equation}

Since ${\mathcal{P}^H}\mathcal{P} + {\beta _1}I + {\beta _3}I$ is a diagonal matrix, $X$  is easily computed by:
\begin{equation}
{X^{k + 1}} \!=\! {{\cal F}^H} \!\! \left[ {\frac{{{{\cal P}^H}Y \!+\! {\cal F}\left( {{\beta _1}\!\left( {{Z^{k + 1}} \!-\! u_Z^k} \right) \!+\! {\beta _3}\!\left( {{B^{k + 1}} \!-\! u_B^k} \right)} \right)}}{{{{\cal P}^H}{\cal P} \!+\! {\beta _1}I \!+\! {\beta _3}I}}} \right] \label{eq35}
\end{equation}

Now that subproblems (\ref{eq24})-(\ref{eq30}) can be solved, we obtain the SPIRiT PMRI reconstruction algorithm, denoted as ADMM-NLR-SPIRiT.




\bibliographystyle{elsarticle-num}

\end{document}